\documentclass[11pt]{article}

\usepackage{amsmath,amssymb}
\usepackage{graphicx}
\usepackage{caption}
\usepackage{color}
\usepackage{dcolumn}
\usepackage{bm}
\usepackage{float}
\usepackage{hyperref}
\usepackage{apacite}
\hypersetup{colorlinks = true, citecolor = black, urlcolor = blue}

\definecolor{background-color}{gray}{0.98}

\usepackage[shortlabels]{enumitem}
\newlist{arguments}{description}{1}
\setlist[arguments]{style=sameline}
\newlist{inlinelist}{enumerate*}{1}
\setlist*[inlinelist,1]{label=(\alph*), itemjoin={{, }}, itemjoin*={{, and }}}

\usepackage{algpseudocode}

\algnewcommand\algorithmicinput{\textit{Input:}}
\algnewcommand\Input{\item[\algorithmicinput]}
\algnewcommand\algorithmicoutput{\textit{Output:}}
\algnewcommand\Output{\item[\algorithmicoutput]}
\newcommand{\algrule}[1][1pt]{\par\vskip.5\baselineskip\hrule height #1\par\vskip.5\baselineskip}

\newcounter{ssaalg}[section]
\makeatletter
\renewcommand\thessaalg{\thechapter.\@arabic\c@ssaalg}%
\def\ext@ssaalg{loa}
\def\fnum@ssaalg{\textsc{Algorithm~\thessaalg}}
\makeatother

\graphicspath{{img_ch1/}{img_chm/}{img_ch2d/}{img/}}

\usepackage{fancyvrb}
\DefineVerbatimEnvironment{Code}{Verbatim}{fontsize=\small,fontshape=sl}
\DefineVerbatimEnvironment{Func}{Verbatim}{fontsize=\small,fontshape=tt}
\DefineVerbatimEnvironment{CodeInput}{Verbatim}{fontsize=\footnotesize, fontshape=sl}
\DefineVerbatimEnvironment{CodeInput}{Verbatim}{fontsize=\scriptsize, fontshape=sl}

\makeatletter
\newcommand\code{\bgroup\@makeother\_\@makeother\~\@makeother\$\@codex}
\def\@codex#1{{\small\ttfamily\hyphenchar\font=-1 #1}\egroup}
\makeatother

\let\proglang=\textsf
\usepackage{xspace}
\def\R{{\normalfont\ttfamily R}\xspace}

\makeatletter
\def\verbatim@font{\small\ttfamily}
\makeatother

\usepackage{euscript}

\newcommand{\rmF}{\mathrm{F}}

\newcommand{\rmT}{\mathrm{T}}

\newcommand{\tN}{\mathsf{N}}

\newcommand{\tP}{\mathsf{P}}

\newcommand{\tR}{\mathsf{R}}
\newcommand{\tS}{\mathsf{S}}
\newcommand{\tT}{\mathsf{T}}

\newcommand{\tX}{\mathsf{X}}

\newcommand{\bfC}{\mathbf{C}}

\newcommand{\bfH}{\mathbf{H}}

\newcommand{\bfM}{\mathbf{M}}

\newcommand{\bfS}{\mathbf{S}}

\newcommand{\bfU}{\mathbf{U}}

\newcommand{\bfX}{\mathbf{X}}

\newcommand{\calD}{\mathcal{D}}

\newcommand{\calH}{\mathcal{H}}

\newcommand{\calM}{\mathcal{M}}

\newcommand{\calT}{\mathcal{T}}

\newcommand{\bt}{\begin{theorem}}
\newcommand{\et}{\end{theorem}}
\newcommand{\bl}{\begin{lemma}}
\newcommand{\el}{\end{lemma}}
\newcommand{\bp}{\begin{proposition}}
\newcommand{\ep}{\end{proposition}}
\newcommand{\bc}{\begin{corollary}}
\newcommand{\ec}{\end{corollary}}

\newcommand{\bd}{\begin{definition}\rm}
\newcommand{\ed}{\end{definition}}
\newcommand{\bex}{\begin{example}\rm}
\newcommand{\eex}{\end{example}}
\newcommand{\br}{\begin{remark}\rm}
\newcommand{\er}{\end{remark}}

\newcommand{\btbh}{\begin{table}[!ht]}
\newcommand{\etb}{\end{table}}
\newcommand{\bfgh}{\begin{figure}[!ht]}
\newcommand{\efg}{\end{figure}}

\newcommand{\bea}{\begin{eqnarray*}}
\newcommand{\eea}{\end{eqnarray*}}
\newcommand{\be}{\begin{eqnarray}}
\newcommand{\ee}{\end{eqnarray}}

\newcommand{\suml}{\sum\limits}

\newcommand{\ve}{\varepsilon}

\newcommand{\lm}{\lambda}
\def\wtilde{\widetilde}

\def\spaceR{\mathbb{R}}
\newcommand\Expect{\mathrm{E}}

\def\last#1{{\underline{#1}}}
\def\first#1{{\mathstrut\overline{#1}}}

\def\sspan{\mathop{\mathrm{span}}}
\def\rank{\mathop{\mathrm{rank}}}

\newcommand{\Arg}{\mathop\mathrm{Arg}}

\makeatletter
\def\adots{\mathinner{\mkern2mu\raise\p@\hbox{.}
\mkern2mu\raise4\p@\hbox{.}\mkern1mu
\raise7\p@\vbox{\kern7\p@\hbox{.}}\mkern1mu}}
\newcommand{\l@abcd}[2]{\hbox to\textwidth{#1\dotfill #2}}
\makeatother

\usepackage[english]{babel}

\usepackage{tikz}
\usepackage{color}
\usepackage{subfig}

\makeatletter
\def\adotsss{
\begin{picture}(6,7)
\put(0,2){\circle*{1}}
\put(3,4){\circle*{1}}
\put(6,6){\circle*{1}}
\end{picture}
}
\makeatother

\makeatletter
\def\vdotss{
\begin{picture}(1,7)
\put(0,1){\circle*{1}}
\put(0,4){\circle*{1}}
\put(0,7){\circle*{1}}
\end{picture}
}
\makeatother

\def\Var{\mathsf{D}}

\def\colspace{\mathop{\mathrm{colspace}}}

\def\Mod{\mathop\mathrm{Mod}}
\def\Arg{\mathop\mathrm{Arg}}

\newtheorem{theorem}{Theorem}
\newtheorem{proposition}{Proposition}
\newtheorem{definition}{Definition}
\newtheorem{remark}{Remark}[section]

\usepackage{fleqn}

\title{Particularities and commonalities of singular spectrum analysis as a method of time series analysis and
signal processing}

\author{Nina Golyandina\thanks{n.golyandina@spbu.ru, St.Petersburg State University, Universitetskaya nab. 7--9. St.Petersburg, 199034, Russia}}

\date{}
\begin{document}
\maketitle

\begin{abstract}
  Singular spectrum analysis (SSA), starting from the second half of the XX century, has been a rapidly developing method of time series analysis. Since it can be called principal component analysis for time series, SSA will  definitely be a standard method in time series analysis and signal processing in the future. Moreover, the problems solved by SSA  are considerably wider than that for principal component analysis. In particular, the problems of frequency estimation, forecasting and missing values imputation can be solved within the framework of SSA. The idea of SSA came from different scientific communities, such as that of researchers in time series analysis (Karhunen-Lo\`{e}ve decomposition), signal processing (low-rank approximation and frequency estimation) and multivariate data analysis (principal component analysis). Also, depending on the area of applications, different viewpoints on the same algorithms, choice of parameters, and methodology as a whole are considered. Thus, the aim of the paper is to describe and compare different viewpoints on SSA and its modifications and extensions to give people from different scientific communities the possibility to be aware of potentially new aspects of the method.
\end{abstract}

\tableofcontents

%\newpage
\section{Introduction}

\subsection{References}
The origin of singular spectrum analysis (SSA) is usually referred to the papers \cite{Broomhead.King1986} and \cite{Fraedrich1986}; although, the algorithm of SSA can be found e.g. in \cite{Colebrook1978}.
SSA became widely known in climatology after publication of \cite{Vautard.Ghil1989,Vautard.etal1992}.
After several years, the book \cite{Elsner.Tsonis1996} summarized the basic information
about SSA existing to that moment.
{In parallel, SSA (named `Caterpillar') was created in Russia; the results were published in \cite{Danilov.Zhigljavsky1997ru} (in Russian). The history of the `Caterpillar' method starts from
\cite{Belonin.etal1971}, where O.~M. Kalinin is indicated as the author of the ideas underlying the method (it is difficult to find the access to this review; therefore, we refer to the book
\cite[Chapter 3, Section 8]{Belonin.etal1982}, where this review is cited together with the algorithm of the first stage of SSA). Another source of the Russian branch was the paper \cite{Buch94}.}

A breakthrough in the theory of SSA was made in the fundamental book \cite{Golyandina.etal2001}, where the theory is presented together with examples.
The next book is \cite{Golyandina.Zhigljavsky2013} in the series Briefs in Statistics; it contains a brief description and some updates from 2001, including description of SSA as a set of filters.
From 2013, a large jump was performed, when SSA became a method for analysis of objects of different dimensions and shapes. Also, the \textsf{R}-package \textsf{Rssa} \cite{Rssa} was developed with a very fast implementation of SSA for different kinds of objects. The proposed structured approach to SSA, its multivariate extensions (MSSA and 2D-SSA) together with algorithms and description of the implementation in \textsf{Rssa} are contained in the recent book  \cite{Golyandina.etal2018}.

Three mentioned monographs of Golyandina and coauthors cover a very wide range of problems solved by SSA; however, they only briefly discuss practical applications of SSA to stationary processes. At the same time, the applications of SSA to stationary time series were developed by the team from UCLA (starting from \cite{Yiou.etal2000}), mostly for climatic data. Some practical applications, in particular, in economics and biomedicine, are considered in the works of H.Hassani, S.Sanei and their coauthors (see, e.g., the book \cite{Sanei.Hassani2015} and the review \cite{Hassani.Thomakos2010}). A separate branch is related to real-world problems in geophysics, where traces in the form of straight lines should be extracted; a preliminary processing is performed by the discrete Fourier transform of
the image rows and then Complex SSA is applied to the Fourier coefficients \cite{Trickett2003,Oropeza2010}.
These branches seem to be developing somewhat independently; therefore, it would be very helpful to enrich one another.

\subsection{Sketch of the algorithm} The SSA algorithm consists of two stages. The first stage is called Decomposition, where
the studied object (e.g. a time series) is transformed into a trajectory matrix (a Hankel matrix) and then the singular value decomposition is applied to
the trajectory matrix to obtain a decomposition into elementary rank-one matrix components.
The second stage, which is called Reconstruction, creates grouped matrix components in a clever way
and transforms the grouped matrix decomposition back to a decomposition of the
initial object by the so-called diagonal averaging.

Since the idea of considering the subseries of one time series as different observations and then applying principal component analysis (PCA) or Karhunen-Lo\`{e}ve transform (KLT) to the obtained sample is straightforward,
Decomposition stage of SSA can be found in many papers; it is hard to
detect which paper was first. For example, the mentioned above papers \cite{Belonin.etal1971, Broomhead.King1986, Fraedrich1986, Vautard.Ghil1989} contain description only of Decomposition stage.
Also, we can cite \cite{Basilevsky.Hum1979} and \cite{Efimov.Galaktionov1983} as references related to the first stage of SSA.
`Diagonal averaging' from the second stage is used in \cite{Colebrook1978, Tufts.etal1982, Cadzow1988, Ghil.Vautard1991, Buch94}.
Nowadays, Reconstruction stage is considered as an essential part of SSA.

Another origin of SSA traces back to properties of Hankel matrices \cite{Gantmacher1959}.
Sometimes, an origin of SSA is drawn from \cite{Prony1795}, where the modelling of the signal in the form of a sum of exponential series was considered; this origin is related to the parameter (frequency) estimation problem.

We suggest to call the method `SSA' if both Decomposition and Reconstruction stages are involved.
The methods based on Decomposition stage only are called subspace-based methods.
Although many subspace-based methods were developed before SSA,  these methods may be called SSA-related.

\subsection{Motivation}
Depending on the area of applications, different points of view on the same algorithms, the choice of parameters and the methodology as a whole are considered. The aim of the present paper is
   to describe and compare different points of view on SSA and its modifications and extensions to
   give people from different scientific communities the possibility to be aware of potentially
   new aspects of the method.
\subsection{Structure}
{Section~\ref{sec:review} contains the description of singular spectrum analysis starting from its algorithm and basic ideas. We briefly discuss SSA from different viewpoints to show that the origins of SSA have connections to a very wide range of known methods. Particular cases and extensions of SSA are described.
The emphasis is placed upon the differences in descriptions of SSA-related methods in various papers.
 Section~\ref{sec:SSAand} contains specific problems of time series analysis and their solution by SSA in comparison
 with other methods.
 In several subsections of Section~\ref{sec:SSAand} we touch the same abilities of SSA as in Section~\ref{sec:review}; however, we look at these abilities from different angles. To minimize intersections between such sections, we provide cross-references.}
 In Section~\ref{Rssa}, we briefly describe implementations of SSA, since in the era of big data,
 the effective implementation of a method is a key point for its use in real-world problems.
Section~\ref{sec:concl} concludes the paper.

\smallskip
This review paper is supported by the monograph \cite{Golyandina.etal2018}, which describes and organizes the algorithms of SSA-related methods and their implementation in the \textsf{R}-package \textsf{Rssa} \cite{Rssa} with numerous examples.
The monograph contains the description of different aspects of SSA itself; in this paper, we put an emphasis on different external viewpoints on SSA and their connection with other methods. To avoid the repetition of figures and to give the readers convenient possibility to
look at the illustrative pictures, we put into footnotes the links to the examples
from the companion website \url{https://ssa-with-r-book.github.io} to the book \cite{Golyandina.etal2018}.

\subsection{Detailed structure}
\subsubsection{A general review of SSA (Section~\ref{sec:review})}
\begin{enumerate}
\item Time series and digital images: common problems (Section~\ref{sec:ts_common})
%\begin{itemize}
%\item Decomposition
%\item Other time-series/image problems
%\end{itemize}
\item Singular spectrum analysis (Section~\ref{sec:ssa_common})
\begin{itemize}
\item An idea of SSA
\item Basic SSA algorithm
\item A view from the Karhunen-Lo\`{e}ve transform
\item A view from stationary processes
\item A view from dynamical systems
\item A view from structured low-rank approximation
\end{itemize}
\item Decomposition (Section~\ref{sec:decomp_common})
\begin{itemize}
\item Separability
\item How to identify the SVD components
\end{itemize}
\item Filtering (Section~\ref{sec:filtering})
\item Modelling (Section~\ref{sec:modelling})
\begin{itemize}
\item Subspace-based approach
\item Signal extraction via projections
\end{itemize}
\item Choice of parameters (Section~\ref{sec:choice_param})
\item Theoretical studies (Section~\ref{sec:theory})
\item General scheme of SSA decompositions (Section~\ref{sec:general})
\item Multivariate/multidimensional extensions (Section~\ref{sec:multi})
\begin{itemize}
\item Multivariate SSA
\item Two-dimensional SSA
\item Shaped SSA
\item Complex SSA
\end{itemize}
\item Modifications of the SVD step (Section~\ref{sec:modif})
\begin{itemize}
\item Use of a priori information
\item Refined decompositions of signals
\item Tensor SSA
\end{itemize}
\end{enumerate}
\subsubsection{SSA and different problems (Section~\ref{sec:SSAand})}
\begin{enumerate}
\item SSA and nonlinearity. Is SSA a linear method? (Section~\ref{sec:linear})
\item SSA and autoregressive processes (Section~\ref{sec:AR})
\item SSA and parameter estimation (Section~\ref{sec:parameters})
\item SSA and structured low-rank approximation (Section~\ref{sec:SLRA})
\item SSA and linear regression (Section~\ref{sec:linear_trend})
\item SSA and filtering (Section~\ref{sec:SSA_filtering})
\item SSA and independent component analysis (Section~\ref{sec:ICA})
\item SSA and empirical mode decomposition, discrete Fourier transform and discrete wavelet transform (Section~\ref{sec:spectral})
\item SSA: model-free method and modelling (Section~\ref{sec:SSA_modelling})
\item SSA: forecasting and gap filling (Section~\ref{sec:forecast})
\item SSA and signal detection: Monte Carlo SSA (Section~\ref{sec:montecarlo})
\item SSA and outliers (Section~\ref{sec:outliers})
\item SSA and a priori/a posteriori information (Section~\ref{sec:apriori})
\item SSA: automatic identification and batch processing (Section~\ref{sec:ident})
\item SSA and machine learning (Section~\ref{sec:ML})
\end{enumerate}
\subsubsection{Implementation of SSA (Section~\ref{Rssa})}
\begin{enumerate}
\item Software and fast implementation (Section~\ref{sec:software})
\item Example of calculations in the \R{}-package \textsf {Rssa} (Section~\ref{sec:ex_rssa})
\end{enumerate}

\section{A general review of SSA}
\label{sec:review}
 \subsection{Time series and digital images: common problems}
\label{sec:ts_common}
\subsubsection{Decomposition}
\begin{figure}[!h]
        \begin{center}
        \includegraphics[width=4in]{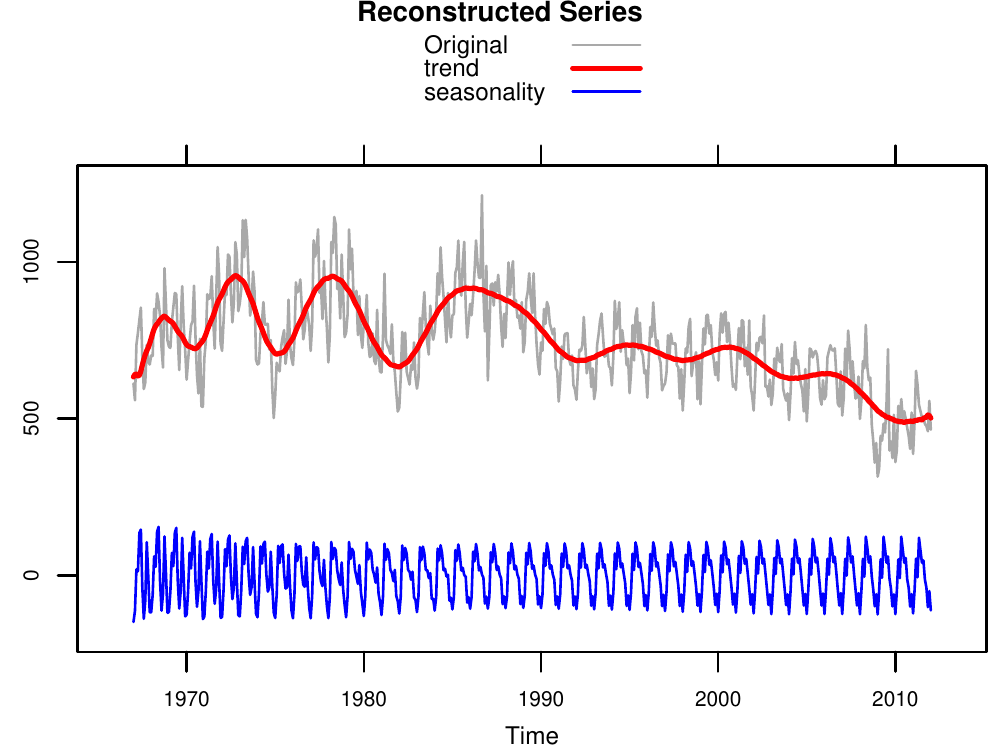}
        \end{center}
        \caption{`MotorVehicle', monthly sales: Decomposition.}
        \label{fig:motor_decomposition}
\end{figure}

Let us observe $\tX = (x_1,\ldots,x_N)$, where $\tX = \tT + \tP + \tN$,  $\tT$ is a trend, $\tP$ contains regular oscillations and $\tN$ is noise.
 The common problem is to construct a decomposition $\tX = \widetilde\tT + \widetilde\tP + \widetilde\tN$, see Fig.~\ref{fig:motor_decomposition}, which allows one
 to estimate the trend $\tT$, the whole signal $\tS=\tT+\tP$, or
 periodic components $\tP$ ($\tP$ can consist of periodic components with different fundamental periods).
 Once the signal is estimated and its structure is detected,
 different signal-based methods such as forecasting can be applied.

 If the time series does not contain a signal, then the problem of filtering can be stated; this problem is reduced again to constructing the decomposition, into low- and high-frequency components.

\begin{figure}[!h]
        \begin{center}
        \includegraphics[height=3.5cm]{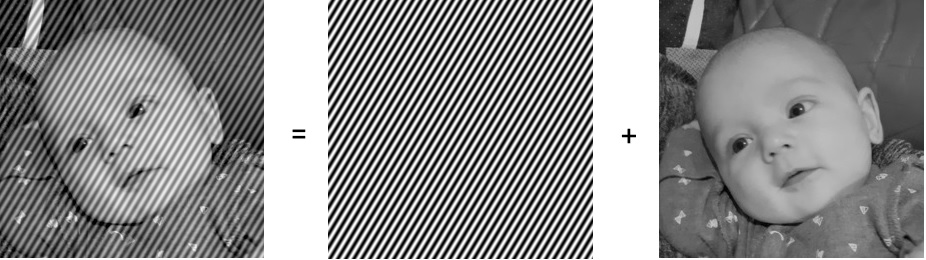}
       \end{center}
        \caption{`Maya': Decomposition.}
        \label{fig:gentlemen_decomposition}
\end{figure}

 Many problems in digital image processing are similar to the problems stated for time series. Let
 \begin{equation*}
 \tX = \left(
 \begin{array}{lll}
 x_{1,1}&\ldots&x_{1,N_2}\\
 \ldots&\ldots&\ldots\\
 x_{N_1,1}&\ldots&x_{N_1,N_2}
 \end{array}
 \right)
\end{equation*}
 be a digital image, which is modelled as a decomposition
 $\tX = \tT + \tP + \tN$ into a pattern, regular oscillations (e.g. a texture)
and noise. Then the problem of estimating the decomposition components arises again.
 See Fig.~\ref{fig:gentlemen_decomposition}, which demonstrates the extraction of a pattern
 (which is obtained after the removal of regular oscillations).

Note that `digital images' is a common name for 2D data, since one of the dimensions may be temporal.
The decomposition problem for multidimensional data is also important in higher dimensions;
these dimensions may again be of different nature.
For example, one can consider  both 3D spatial data and  2D data with the third temporal dimension as 3D data; and so on. We will call data with $n$ dimensions $n$D data.

\subsubsection{Other time-series/image problems}

The results of decomposition (in particular, of signal extraction) allow one to solve many problems for objects
with different dimensions and shapes; among these problems are
\begin{itemize}
\item
trend/tendency extraction;
\item
smoothing/filtering;
\item
noise reduction;
\item
extraction of periodic components (including seasonal adjustment);
\item
frequency estimation;
\item
construction of a model and parameter estimation;
\item
forecasting/prediction (of the extracted signal);
\item
missing data imputation;
\item
change-point detection.
\end{itemize}
Thus, the decomposition serves as a starting point for solving many important problems.

\subsection{Singular spectrum analysis}
\label{sec:ssa_common}
In the paper, we will consider the application of singular spectrum analysis (SSA) for most of the problems
mentioned above.
The name `SSA' can be used in both a narrow and a broad sense.
One can talk about SSA in a narrow sense meaning the algorithm of decomposition of an object
into a sum of identifiable components.
SSA in a broad sense consists of methods that use the results of the SSA decomposition.
Let us start with the description of SSA as a decomposition method.

\subsubsection{An idea of SSA: to create samples of structure}
In multivariate data, there are many observations, which are sampled from the same probability distribution.
In data like time series or digital images, we have only one object and therefore need to create many samples of the object's structure to detect it.
This can be done by a moving procedure.

Let us start with time series. Denote $\tX = (x_1,\ldots,x_N)$ the time series of length $N$ and choose a window length $L$, $1<L<N$; then consider $K=N-L+1$ vectors constructing from the moving subseries of length $L$:
$X_1 =(x_1,\ldots,x_{L})^{\rmT}$, $X_2=(x_2,\ldots,x_{L+1})^{\rmT}$, $\ldots$ (the procedure looks like a caterpillar is moving;
this step led to one of the method names `Caterpillar').

For digital images
 $$\tX = \left(
 \begin{array}{lll}
 x_{1,1}&\ldots&x_{1,N_2}\\
 \ldots&\ldots&\ldots\\
 x_{N_1,1}&\ldots&x_{N_1,N_2}
 \end{array}
 \right),
$$
the size of the moving 2D windows is $L_1\times L_2$; the moving procedure is performed in two directions,
from left to right and from top to bottom:
$$\tX^{(L_1,L_2)}_{k,m} = \left(
 \begin{array}{lll}
 x_{k,m}&\ldots&x_{k,m+L_2-1}\\
 \ldots&\ldots&\ldots\\
 x_{k+L_1-1,m}&\ldots&x_{k+L_1-1,m+L_2-1}
 \end{array}
 \right)
,$$
$k=1,\ldots, K_1$ and $m=1,\ldots, K_2$, where $K_1 = N_1-L_1+1$, $K_2= N_2-L_2+1$

\subsubsection{An idea of SSA: to find a common structure using the SVD}

Using the obtained set of sub-objects of the given object, we should find a common structure.
There is a standard statistical method for this, principal component analysis (PCA), which is
applied to data in matrix form.
Hence, the moving sub-objects are transformed into vectors, which are then stacked in a matrix called
\emph{trajectory matrix}.

Note that PCA is applied to matrices, whose rows and columns have different nature
since if the rows correspond to variables, then the columns contain observations. In the case of SSA, the rows and columns of the trajectory matrix have typically the same structure; for time series, both the rows and the columns are subseries of the same time series.
Whereas PCA consists of centering or standardization of the variables (the rows of the data matrix) and
then applying the SVD to the transformed matrix, SSA usually omits the centering/standardization.
The version of SSA which directly applies the SVD to the trajectory matrix is called Basic SSA.

Due to the approximation properties, the SVD allows one to extract the signal (i.e., to remove noise).
Due to the biorthogonality of the SVD, the separation of the signal components (e.g. trend and periodic
components) one from another is possible. {(We refer the reader to \cite[Chapter 4]{Golyandina.etal2001},
where the properties of the SVD, which are helpful for understanding SSA, are gathered.)}

For the separation of the signal components one from another, the approximation properties are not necessary.
Therefore, other techniques (instead of the SVD) can be applied after the signal extraction.
One can see an analogy with the factor analysis of multivariate data, where latent variables should be extracted;
first, a number of initial factors are found (e.g. using PCA) and then the rotation of these variables is used for better interpretability. The rotated factors do not have approximation properties.

\subsubsection{Basic SSA algorithm}
\label{sec:basic_ssa}
Here we introduce the algorithm for SSA decomposition of time series.
The general form of the algorithm is given in \cite{Golyandina.etal2018} and Section~\ref{sec:general}.

Let $\tX = (x_1,\dots,x_{N})$ be a time series of length $N$.

\paragraph{Decomposition stage} (Parameter: window length $L$;\quad$1<L<N$)
\begin{enumerate}
\item {Embedding}\\
The trajectory matrix is constructed by means of the embedding operator $\calT$, which maps
a time series to an $L\times K$ Hankel matrix ($K=N-L+1$) as follows:
\begin{equation}
\label{eq:traj}
\calT(\tX)=\bfX = \left(\!{\renewcommand{\arraystretch}{1.1}
\begin{array}{@{\ }l@{\;\;}l@{\;\;}l@{\;\;}l@{\;}}
x_1      & x_2 & \dots     & x_{K}     \\
x_2      & \;\adotsss  & \;\adotsss    & x_{K+1}    \\
\; \vdotss   & \;\adotsss   & \ \:\vdotss    & \; \vdotss   \\
x_{L}  & x_{L+1} & \dots & x_{N}
\end{array}
}\right).
\end{equation}
\item {Singular Value Decomposition (SVD)}\\ \smallskip
The SVD is given by $\bfX= \suml_{m=1}^d  \sqrt{\lm_m}{U_m}{V_m^\rmT}$,  where $\{U_m\}_{m=1}^d$ and $\{V_m\}_{m=1}^d$ are orthonormal systems of left and right singular vectors of $\bfX$, respectively; $\lm_1\ge\lm_2\ge\ldots\lm_d >0$ are squared singular values.
\end{enumerate}

\paragraph{Reconstruction stage} (Parameter: {way of grouping} $\{1,\ldots,d\}=\bigcup_{j=1}^c I_j$)
\begin{enumerate}
\item {Grouping} \\
The SVD components are grouped: $\bfX = \bfX_{I_1} + \ldots + \bfX_{I_c}$, where
$\bfX_I = \suml_{m\in I} \sqrt{\lm_m}{U_m}{V_m^\rmT}$.

\item {Diagonal averaging}\\
Each matrix $\bfX_I$ is transferred to the nearest Hankel matrix $\wtilde \bfX_I$ by hankelization and then
$\wtilde \bfX_I$ is transformed into a time series as $\wtilde\tX^{(I)}=\calT^{-1}(\wtilde \bfX_I)$.
\end{enumerate}

Thus, the output of the SSA algorithm is the decomposition $\tX = \wtilde\tX^{(I_1)} + \ldots + \wtilde\tX^{(I_c)}$, which corresponds to the structure of the original time series $\tX$
if a proper grouping was used.

\subsubsection{Comments to Basic SSA}
\label{sec:comments}
\paragraph{Singular Value Decomposition (SVD)}
The SVD
$\bfX= \suml_{m=1}^d  \sqrt{\lm_m}{U_m}{V_m^\rmT}$ is the biorthogonal decomposition into a sum of rank-one matrices, $d=\rank{\bfX}$. By the Eckart-Young theorem \cite{Eckart.Young1936}, it
provides the best low-rank approximation in the Frobenius norm.

The computation of the SVD can be reduced to finding eigenvectors $\{U_m\}_{m=1}^d$ and eigenvalues $\{\lambda_m\}_{m=1}^d$ of the matrix $\bfX \bfX^\rmT$, which is sometimes called lag-covariance matrix (despite of centering is not applied); $V_m = \bfX^\rmT U_m/\sqrt{\lambda_m}$.
The collection $(\sqrt{\lm_m},{U_m},{V_m})$ is called the $m$th eigentriple (ET).

If the trajectory matrix $\bfX$ is considered as a (transposed) data matrix with $K$ cases and $L$ variables,
the SVD of $\bfX$ is closely related to principal component analysis. If rows of $\bfX$ are centered,
the SVD exactly corresponds to PCA. Therefore, according to statistical terminology, $V_m$ and $\sqrt{\lambda_m}V_m$ can be called \emph{factor scores} and \emph{principal components}, respectively.

\paragraph{Grouping} Let $\tX=\tX^{(1)}+\tX^{(2)}$ and therefore the following equality be valid for
the trajectory matrices: $\bfX=\bfX^{(1)}+\bfX^{(2)}$. In the case, when there exists a grouping
$\{1,\ldots,d\}=I_1\cup I_2$ such that $\bfX^{(1)} = \bfX_{I_1}$ and $\bfX^{(2)} = \bfX_{I_2}$
(this is called `separability', see Section~\ref{sec:separ}), the eigenvectors $\{U_m\}_{m\in I_j}$ form a basis of
$\colspace(\bfX^{(j)})$; therefore, $U_m$ for $m\in I_j$ repeats the behavior of $\tX^{(j)}$  (recall that the columns of $\bfX^{(j)}$ are subseries of $\tX^{(j)}$). Thus, we can say
that  an eigenvector repeats the behaviour of a component that produces this eigenvector.
Thereby, we can identify slowly-varying $U_m$ and gather them into the trend group; then identify
regular oscillations and gather them into the periodicity group, and so on.

The elementary grouping, when each group consists of one SVD component (one eigentriple), produces the so-called
\emph{elementary reconstructed time series}. Since the grouping and the diagonal averaging are linear operations,
it does not matter what operation is performed first, the grouping and then the diagonal averaging or
vice versa. Thereby, inspection of elementary reconstructed time series is helpful for performing grouping,
since the grouping is simply the summation of these time series.

There is a particular case, when a noisy signal is observed and the signal should be extracted.
Formally, this is the case of two groups,
signal and noise ones. Since we need only the signal group, the grouping is reduced
to the choice of one signal group $I$. Due to approximation properties of the SVD,
the group $I$ corresponds to signal usually consisting of $r$ leading components, that is, $I=\{1,\ldots,r\}$. Thus,
the way of grouping is reduced to the choice of $r$.

\subsubsection{A view from the Karhunen-Lo\`{e}ve transform}
For a random process $\xi_t$, $t\in [0,T]$, the structure is contained in the autocovariance function $K(s,t) = \Expect (\xi_s-\Expect\xi_s)(\xi_t-\Expect\xi_t)$;
and the Karhunen-Lo\`{e}ve transform (KLT) based on the eigendecomposition of $K(t,s)$ is considered to express this structure in the form of a decomposition.
Originally, the Karhunen-Lo\`{e}ve decomposition of $\xi_t$ is the decomposition into an infinite sum of white noises $\ve_k$, $k=1,2,\ldots$, with coefficients obtained from the eigenfunctions $u_k(t)$ of the autocovariance function: $\xi_t = \sum_k u_k(t) \ve_k $.

However, for discrete time and a finite time series length, the KLT of $(\xi_1,\ldots,\xi_N)$ in fact coincides with PCA of the multivariate random variable $\bm\xi = (\xi_1,\ldots,\xi_N)^\rmT$. Hence, the Karhunen-Lo\`{e}ve transform is the decomposition into a finite sum of white noises $\ve_k$, $k=1,\ldots,N$, with coefficients obtained from the eigenvectors $U_k$ of the variance-covariance matrix of $\bm\xi$: $\bm\xi = \sum_{k=1}^N U_k \ve_k$.
For a time series $(x_1,\ldots,x_N)$ of length $N$, which can be considered as a realization of $(\xi_1,\ldots,\xi_N)$,
the empirical KLT is constructed on the basis of eigenvectors of the sample covariance matrix.
If the time series is centered or, alternatively, centering is  applied to the rows of the trajectory matrix $\bfX$,
then $\bfX\bfX^\rmT$ can be considered as an estimate of the (auto)covariance matrix.

The question is how to construct the sample version of $\ve_k$, $k=1,\ldots,N$.
In PCA, we have many samples of $(\xi_1,\ldots,\xi_N)^\rmT$. Thus, the sample version of $\ve_k$ is the $k$th
principal component of the multivariate data.
For time series, we have only one realization of length $N$. Constructing the trajectory matrix $\bfX$ is the way to create $K=N-L+1\ $ $L$-dimensional samples. Then $\sqrt{\lambda_m}V_m\in \spaceR^K$ can be considered as the sample version of $\ve_m$,
$m=1,\ldots,\min(L,K)$.

Thus, up to the above assumptions, the empirical KLT formally corresponds to Decomposition stage of the SSA algorithm \cite{Basilevsky.Hum1979}.
People, who studied the theory of random processes and then looked at the SSA algorithm, sometimes say that it is nothing new, just the KLT. However, note that Reconstruction stage is not included in the KLT decomposition.
Also, the point of view on SSA from random process theory leaves a mark on the methodology of applications of SSA (see, e.g., \cite{Khan.Poskitt2013}).

\subsubsection{A view from stationary processes}
\label{sec:stationary}
Researchers who deal with stationary random processes look at SSA in a different way.
For (weakly) stationary processes, the autocovariance function $K(s,t)$ depends on $|t-s|$.
Therefore, one of the main characteristics of a stationary random process $\xi_t$, $t\ge 0$, with mean $\mu$ is its autocovariance function $C(|s-t|) = K(s,t) = \Expect (\xi_s-\mu)(\xi_t-\mu)$ of one variable. For discrete time and
a sequence $\xi_1,\ldots,\xi_N$, this means that the $N\times N$ autocovariance matrix $\{\Expect (\xi_i-\mu)(\xi_j-\mu)\}_{i,j=1}^N$ is Toeplitz. However, in practice, we cannot accurately estimate covariances with large lags, since there are only one pair of observations with lag $N-1$, two pairs of observations with lag $N-2$, and so on.
In SSA, the autocovariances are estimated up to the lag $L$; then not smaller than $K=N-L+1$ pairs of observations can be used for estimating these autocovariances.
Since autocovariances are proportional to autocorrelations for stationary processes,
it does not matter which of them is considered.

In Basic SSA, the lag-covariance matrix $L\times L$ has the form $\bfC = \bfX\bfX^\rmT/K$, which is close to a Toeplitz matrix if the time series is stationary (we assume that the time series is centered); however, this is not exactly a Toeplitz matrix.
 In Toeplitz SSA, the lag-covariance matrix is estimated in a way to obtain exactly a Toeplitz matrix.
 The conventional estimation $\widetilde{C}=\{\widetilde{c}_{ij}\}_{i,j = 1}^{L}$ of the autocovariance matrix is
 $\widetilde{c}_{ij} = \frac{1}{N-|j-i|} \sum\limits_{k=1}^{N-|j-i|} x_k x_{k+|j-i|}$.
In both Basic and Toeplitz SSA, the decomposition step can be expressed as
$\bfX=\sum_m P_m(\bfX^\rmT P_m)^\rmT$. In Basic SSA, $P_m=U_m$ are the eigenvectors of $\bfC$, whereas
in Toeplitz SSA, $P_m$ are the eigenvectors of $\widetilde{C}$.

Since the intention of Toeplitz SSA is a better estimation of the autocovariances,
small enough window lengths are usually chosen to obtain stable estimates of the autocorrelations, which should be
estimated for lags that are not larger than the chosen window length.

Note that in \cite{Vautard.Ghil1989},
the Toeplitz version is called VG according to the author's names, while the basic version
with the SVD decomposition is called BK, since it was suggested in \cite{Broomhead.King1986}.

It is important to note that the only form of the decomposition step which is invariant with respect to
transposition of the trajectory matrix (that corresponds to the change $L'=K$ and $K' = L$)
is the SVD. In particular, for Toeplitz SSA the decompositions for $L$, $K$ and $L'=K$, $K' = L$
are different.

It seems that this view from stationary processes limits the application of SSA, since Toeplitz estimates of autocorrelation matrices have sense for analyzing stationary time series only, while the range of problems solved by SSA is much wider.
Since the VG (Toeplitz) version was proposed for the analysis of climatic data as a default option,
this version is still considered as the main one in many applications; probably, sometimes the Toeplitz version
is used just through habit. However, it should be noted that
the Toeplitz version of SSA is linked to spectral estimation; in such a case, this version is appropriate.

\subsubsection{A view from dynamical systems}
The origin of SSA conventionally refers to the papers devoted to dynamical systems
\cite{Broomhead.King1986} and \cite{Fraedrich1986}, where the problem is based on the Takens embedding theorem
and the method is called singular system analysis.

It seems that the connection between SSA and the Takens theorem is mostly historical.
{Hence, we omit the description of this `embedology' approach.
More details can be found in \cite{Sauer.etal1991}. A discussion about the reliability of the Broomhead\&King approach as a method of estimating the embedding dimension can be found in \cite{Mees.etal1987} and the subsequent comments.}
Nevertheless, the terminology in SSA is partly taken from the theory of dynamical systems. The dynamical system approach introduced the term `embedding' for the first step of the SSA algorithm; also
 the term `trajectory matrix' is due to interpretation
of columns of the trajectory matrix as a sequence of vectors (a trajectory) in a multidimensional space.

\subsubsection{A view from structured low-rank approximation (SLRA)}
It is well-known that rank-deficient Hankel matrices correspond to
time series of a certain structure, see, e.g., \cite{Gantmacher1959}.
{A particular case of the SSA algorithm, when extracting the signal is of interest (Section~\ref{sec:signal}), practically coincides
with one step of the iterative algorithm suggested in
\cite{Cadzow1988} and later named as Cadzow iterations.}

The matrix called `trajectory' in SSA, was called `enhanced matrix' in the Cadzow's paper and
subsequent papers (e.g. \cite{Hua1992} constructs the enhanced matrix in the 2D case for
2D frequency estimation).

{More details related to SLRA can be found in Section~\ref{sec:SLRA}.}

\subsection{Decomposition}
\label{sec:decomp_common})
Here we consider the problem of decomposition of time series into a sum of identifiable
components such as a trend, periodic components and noise.
The first question is whether SSA allows one to find such a decomposition; then, if the answer is `yes', the problem is: how to identify the
time series components with the help of information about the SVD components. Also, the question about the choice
of window length $L$, which leads to a better decomposition, arises. We start with the first question.

\subsubsection{Separability}
\label{sec:separ}
The (approximate) separability  of time series components that are of interest in the study
is needed to perform a grouping of the SVD components
to (approximately) extract these components.

Let $\tX=\tX^{(1)}+\tX^{(2)}$ be an observed time series and $\bfX=\bfX^{(1)}+\bfX^{(2)}$ be the corresponding equality for their trajectory matrices.
The SVD step of Basic SSA provides an expansion $\bfX = \bfX_1 + \ldots + \bfX_d =\sum_{m=1}^d \sqrt{\lm_m}{U_m}{V_m^\rmT}$.
Separability means that the set of eigentriples  $(\sqrt{\lm_{m}},{U_{m}},{V_{m}})$ in the decomposition of the sum $\bfX^{(1)}+\bfX^{(2)}$ is equal to the union of the  eigentriples $(\sqrt{\lm_{m,j}},{U_{m,j}},{V_{m,j}})$ produced by each time series $\bfX^{(j)}$, $j=1,2$. If this is true, the grouping that separates the time series $\tX^{(1)}$ and $\tX^{(2)}$ exists and it is sufficient to identify the corresponding eigentriples. When the proper grouping is fixed, the diagonal averaging step provides the
decomposition of the original time series.

There are two versions of separability related to non-uniqueness of the SVD which is caused by multiple eigenvalues.
\emph{Weak separability} means that there exists an SVD decomposition of the trajectory matrix $\bfX$, which allows one to group its components and thereby to
gather $\tX^{(1)}$ and $\tX^{(2)}$. \emph{Strong separability} means that any SVD allows the proper grouping.
In practice, it is necessary to have the strong separability, since the used numerical method constructs some singular value decomposition, and therefore the chance that it will coincide with the separating one, is vanishing in the absence of strong separability.

\begin{proposition}
    Let $L$ be fixed. Two time series $\tX^{(1)}$ and $\tX^{(2)}$ are   \emph{weakly separable}, if their column trajectory spaces are orthogonal
    and the same is valid for their row trajectory spaces,
    that is,
    $(\bfX^{(1)})^\rmT \bfX^{(2)}=\bf0_{K,K}$ and $\bfX^{(1)} (\bfX^{(2)})^\rmT
    =\bf0_{L,L}$.

    Two time series $\tX^{(1)}$ and $\tX^{(2)}$ are \emph{strongly
        separable}, if they are weakly separable and the sets of singular values
    of their $L$-trajectory matrices are disjoint, that is, $\lambda_{k,1} \neq
    \lambda_{j,2}$ for any $k$ and $j$.
    \end{proposition}

Thus, the condition for (approximate) weak separability is the (approximate) orthogonality
of subseries of length $L$ and the same for subseries of length $K$.
 For example  \cite[Section 6.1]{Golyandina.etal2001}, the time series $x^{(j)}_n=A_j\cos(2\pi\omega_j n+\phi_j)$,  $n=1,\ldots,N$, $j=1,2$,
are asymptotically weakly separable for any $0<\omega_1\neq\omega_2<0.5$ as $N$ tends to infinity.
For exact weak separability, the condition that $L\omega_j$ and $K\omega_j$
are integers should be fulfilled.
For strong separability, we additionally need $A_1 \neq A_2$, since it is easily to find that
 $\|\bfX^{(j)}\|_\rmF^2 = A_j^2LK/2$ and therefore $\lambda_{m,j} = A_j^2LK/4$, $m=1,2$, for $j=1,2$.

Examples of approximately separable time series are: trend and oscillations; slowly-varying  components and noise;
sinusoids with different periods; seasonality and noise.

Theoretically, the approximate separability is a consequence of the asymptotic
separability as $N\rightarrow \infty$. Therefore, the accuracy of decomposition
depends on the convergence rate as $N\rightarrow \infty$.
For example, for separability of sine waves, the convergence rate is
$C/\min(L,K)$, where $C\sim 1/|\omega_1 - \omega_2|$, see \cite[Example 6.7]{Golyandina.etal2001}.
Therefore, for weak separability of sine waves, the choice $L\approx N/2$ is recommended.
Also, a worse separability occurs for sine waves with close frequencies.

Fragments 2.4.1--2.4.3\footnote{\url{https://ssa-with-r-book.github.io/01-chapter2-part1.html\#fragment-241-noisy-sum-of-three-sinusoids-iterative-o-ssa}}
of \cite{Golyandina.etal2018}
 demonstrate the problem of the lack of weak separability for an artificial example and the way of improving the separability by means of a modification of SSA called Iterative Oblique SSA \cite[Section 2.4]{Golyandina.etal2018}.
 Fragment 2.5.1\footnote{\url{https://ssa-with-r-book.github.io/01-chapter2-part1.html\#fragment-251-separation-of-two-sine-waves-with-equal-amplitudes}} shows the problem of the lack of strong separability, also for an artificial example.
 A real-world example, which explains why we need separability, can be found in Section~\ref{sec:sep_ex}.

\subsubsection{How to identify the SVD components}
\label{sec:ident_rules}
The most sophisticated step of SSA is the way of grouping of elementary reconstructed components (RCs).
In this section we will demonstrate how it can be done in an interactive way.

Let us enumerate the main approaches (see \cite[Section 1.6]{Golyandina.etal2001}
and the example in Section~\ref{sec:decomp_ex}):
\begin{enumerate}
\item
To construct the trend group, choose the eigentriples with slowly varying
eigenvectors. The same can be done on the basis of factor vectors or elementary
reconstructed components.
\item
To extract the periodicity with period $T$, find pairs of components similar to
sine/cosine with periods $T/k$, $k=1,\ldots,[(T-1)/2]$ and one saw-tooth component,
which corresponds to the period 2 if $T$ is even.
The mentioned pairs of sine/cosine can be detected in 2D scatterplots of sequential eigenvectors.
\item
To group the components, look at
the matrix of weighted correlations between the elementary
reconstructed components called $w$-correlation matrix. The elementary reconstructed components with strong
$w$-correlation should be put into the same group. In particular, noise produces
correlated components; a sine wave with period larger than 2 produces two $w$-correlated elementary reconstructed time series. Note that the name `correlation' can be a bit ambiguous here, since the vectors are not centered;
that is, the measure called $w$-correlation is, in fact, the cosine of the angle between the vectors.
\end{enumerate}

The enumerated properties can be formalized to obtain methods for automatic identification.
The approaches to automatic identification are considered in Section~\ref{sec:ident} in more detail.

A special task is the choice of components related to the signal. To extract the signal, we should choose a number
of the leading components $r$. If the signal is used for e.g. forecasting, $r$ can be chosen by the minimization of the forecasting errors for historical data.

\subsubsection{Example of identification and decomposition}
\label{sec:decomp_ex}
Let us demonstrate how to visually identify the SSA components
and to obtain the SSA decomposition, by a simple example. In the 1D scatterplots (Fig.~\ref{fig:fort_1d}),
one can find a slowly varying component (ET 1), whereas in the 2D scatterplots (Fig.~\ref{fig:fort_2d}) regular polygons
say about pairs of sine-wave components (ET 2--3 for $T=12$, ET 4--5 for $T=4$, ET 6--7 for $T=6$, ET 8--9 for $T=2.4$ and ET 10--11 for $T=3$). In these figures, $U_i(k)$ denotes the $k$th coordinate
of the $i$th eigenvector obtained in the SVD step of SSA.  The eigenvector numbers are indicated at the captions
of the graphs.  Fig.~\ref{fig:fort_wcor} with the depicted $w$-correlations provides a guess for grouping, since
strongly correlated components should be included in the same group (the black color shows
correlations close to 1; the white color corresponds to zero correlations).

The resultant decomposition into the trend, the seasonality and noise is depicted in Fig.~\ref{fig:fort_decomp}.
This example is performed by the code of Fragments 2.1.1--2.1.3\footnote{\url{https://ssa-with-r-book.github.io/01-chapter2-part1.html\#fragments-211-australian-wines-input-and-212-fort-reconstruction}}
of \cite{Golyandina.etal2018}.

\bfgh
        \begin{center}
        \includegraphics[width=4in]{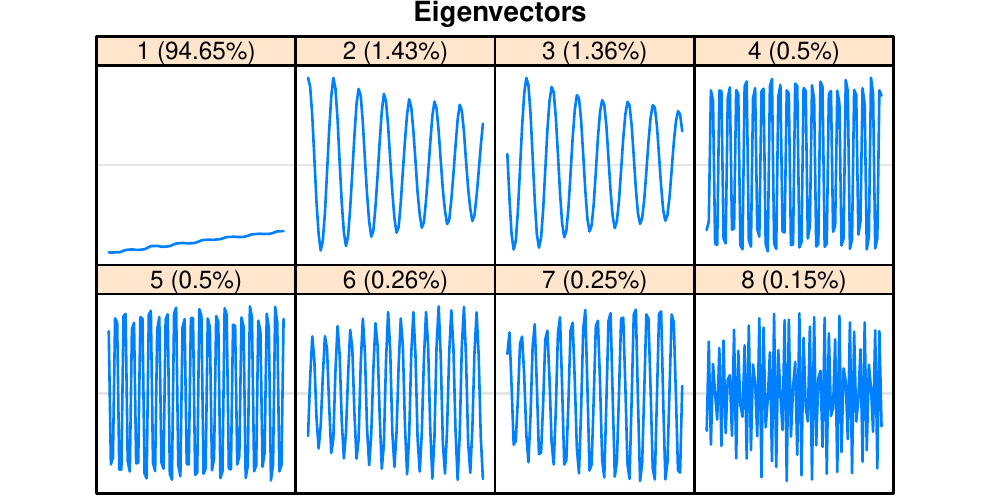}
        \end{center}
        \caption{`Fortified wines', $L=84$: 1D graphs of eigenvectors $(k, U_i(k))$, $k=1,\ldots,L$.}
        \label{fig:fort_1d}
 \efg

 \vspace{-0.4cm}
 \bfgh
        \begin{center}
        \includegraphics[width=4in]{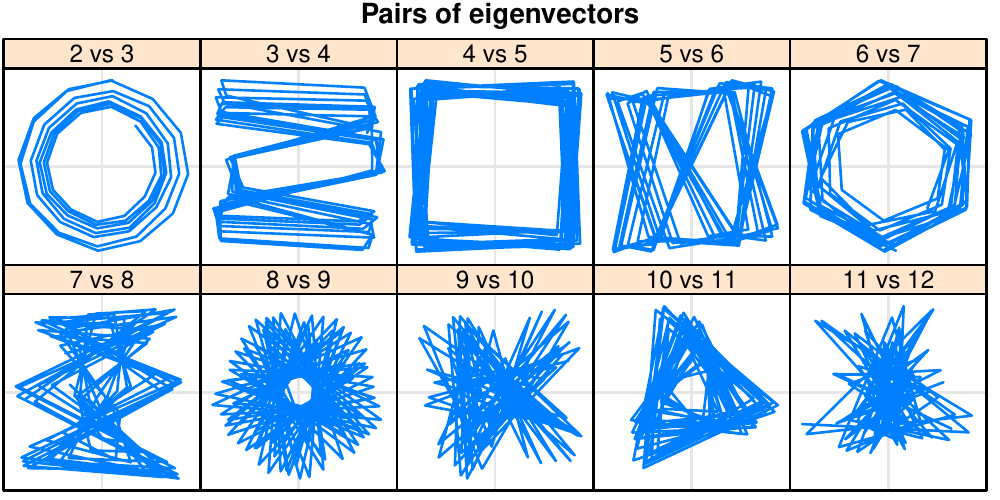}
        \end{center}
        \caption{`Fortified wines', $L=84$: 2D scatterplots of eigenvectors $(U_i(k), U_{i+1}(k))$, $k=1,\ldots,L$.}
        \label{fig:fort_2d}
 \efg

 \vspace{-0.4cm}
\bfgh
        \begin{center}
        \includegraphics[width=4in]{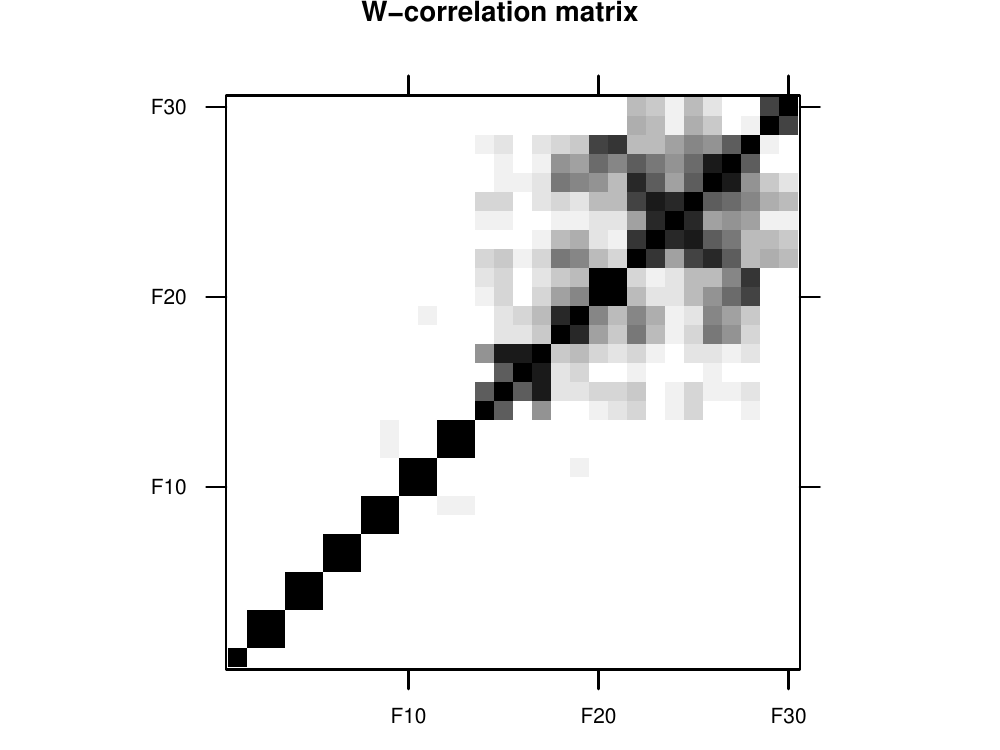}
        \end{center}
        \caption{`Fortified wines', $L=84$: $w$-correlations between elementary RCs.}
        \label{fig:fort_wcor}
\efg

\vspace{-0.4cm}
\bfgh
    \centering
    \includegraphics[width=4in]{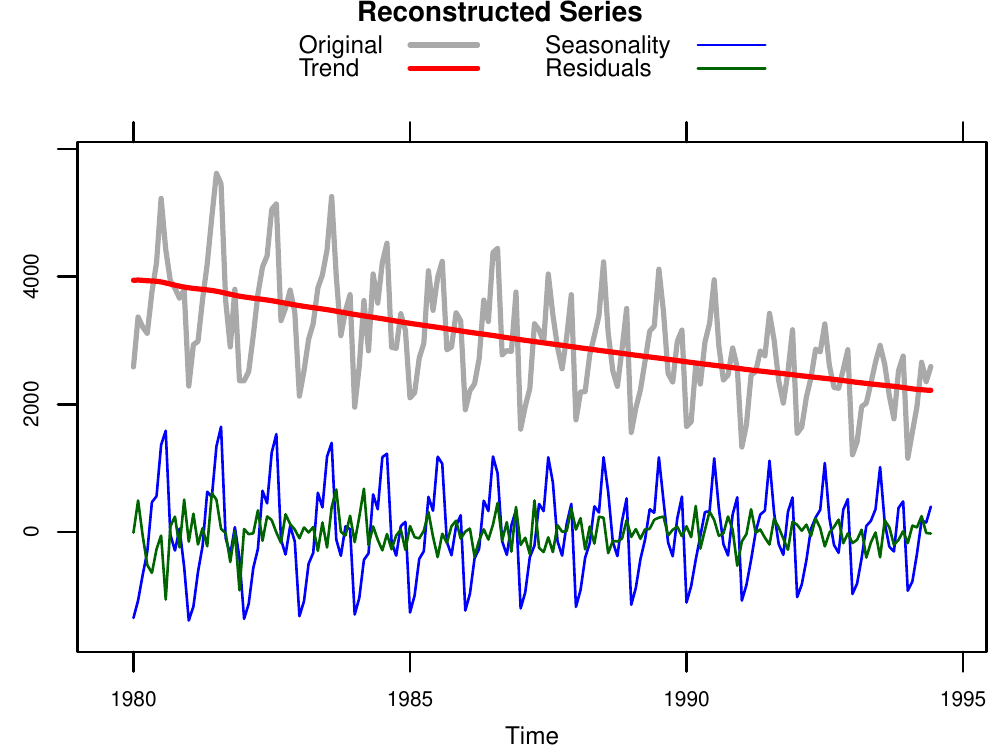}
    %\vspace*{-0.5cm}
    \caption{`Fortified wines', $L=84$: decomposition for groups ET1, ET2--11 and ET12--84.}
    \label{fig:fort_decomp}
\efg

\clearpage
\subsubsection{Example of problems with separability}
\label{sec:sep_ex}
In Section~\ref{sec:decomp_ex}, we considered an example with simple trend and periodic components.
That example corresponds to a good separability between trend, seasonal and noise components.
However, for trends of complex form which are common in time series analysis, there is a big chance that  trend components of the SSA decomposition can be mixed with seasonal components. Below we consider a short time series and demonstrate the problem of the lack of separability.

Let us analyze the first five years of the time series `MotorVehicle' (Fig.~\ref{fig:motor_decomposition}).
The first decomposition is performed by Basic SSA with $L=24$. Figure~\ref{fig:sep_bad} depicts 11 elementary reconstructed series, i.e., the time series, which are obtained with the grouping $\{1,\ldots,d\} = \bigcup_{i=1}^d\{i\}$.
These time series are helpful for the grouping procedure, since the reconstruction by eigentriples from a group
$I$ is just the sum of the elementary reconstructed components with numbers from $I$.

One can see that the trend is contained in slowly-varying RC~1 and partly in RC~2, 8 and 9; the latter elementary reconstructed time series contain a mixture of the trend and seasonality. This means that the reconstruction of the trend with ET1 is insufficient, while that with, say, ET1,2 will contain seasonality (see Figure~\ref{fig:sep_bad_rec}).

{
Thus, if there is no separability of the trend from the residual, then it is impossible to extract an accurate trend. This example explains, why it is very important to try for separability by different means, such as
the choice of the window length, sequential SSA, nested modifications of Basic SSA (e.g., Iterative Oblique SSA,
SSA with derivatives; see Section~\ref{sec:nested}).

Let us apply SSA with derivatives to the group of the 11 leading ETs.
SSA with derivatives changes the chosen components
(but does not change their sum) in such a way to change their contribution and separate them
in the case of the lack of strong separability; the order of the components can be changed
and the trend components are typically the last in the chosen group.
It is clearly seen in Figure~\ref{fig:sep_good} that the trend components are RC~9--11.
Figure~\ref{fig:sep_good_rec} shows the extracted accurate trend.
}

\bfgh
        \begin{center}
        \includegraphics[width=5in]{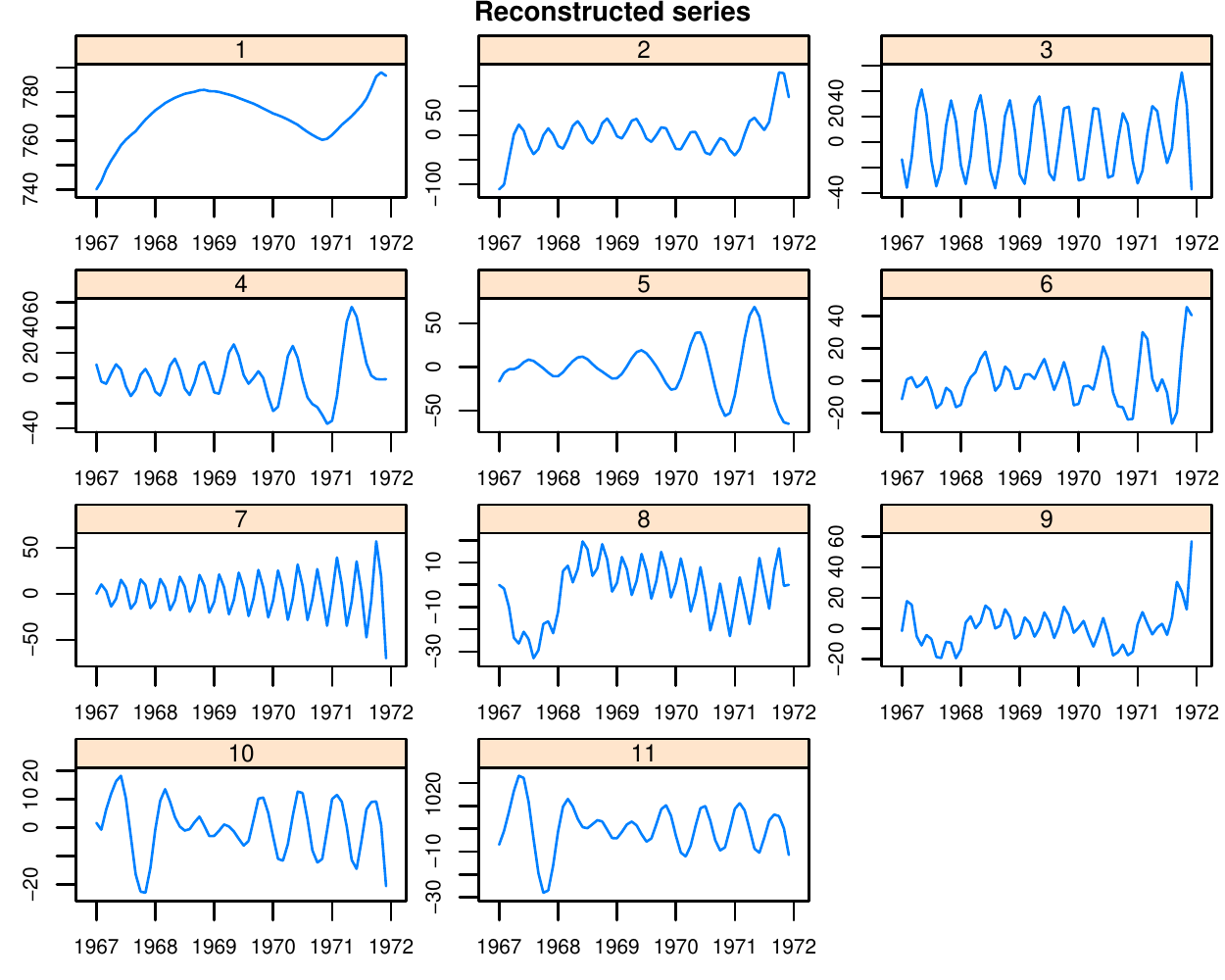}
        \end{center}
        \caption{`MotorVehicle' (5 years), SSA with $L=24$: elementary RCs; poor separability.}
        \label{fig:sep_bad}
\efg

 \vspace{-0.4cm}
\bfgh
        \begin{center}
        \includegraphics[width=4in]{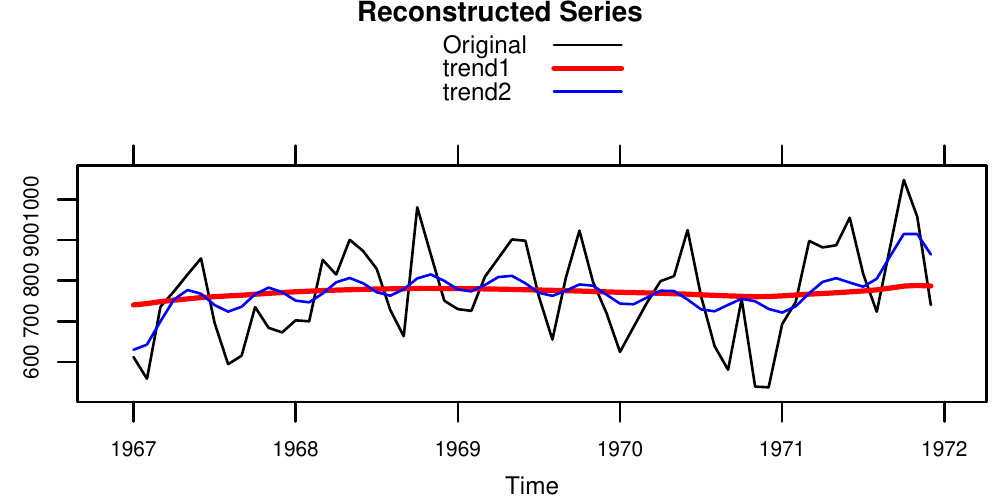}
        \end{center}
        \caption{`MotorVehicle' (5 years), SSA with $L=24$: two trend reconstructions, ET1 and ET1--2; poor separability.}
        \label{fig:sep_bad_rec}
\efg

 \vspace{-0.4cm}
\bfgh
        \begin{center}
        \includegraphics[width=5in]{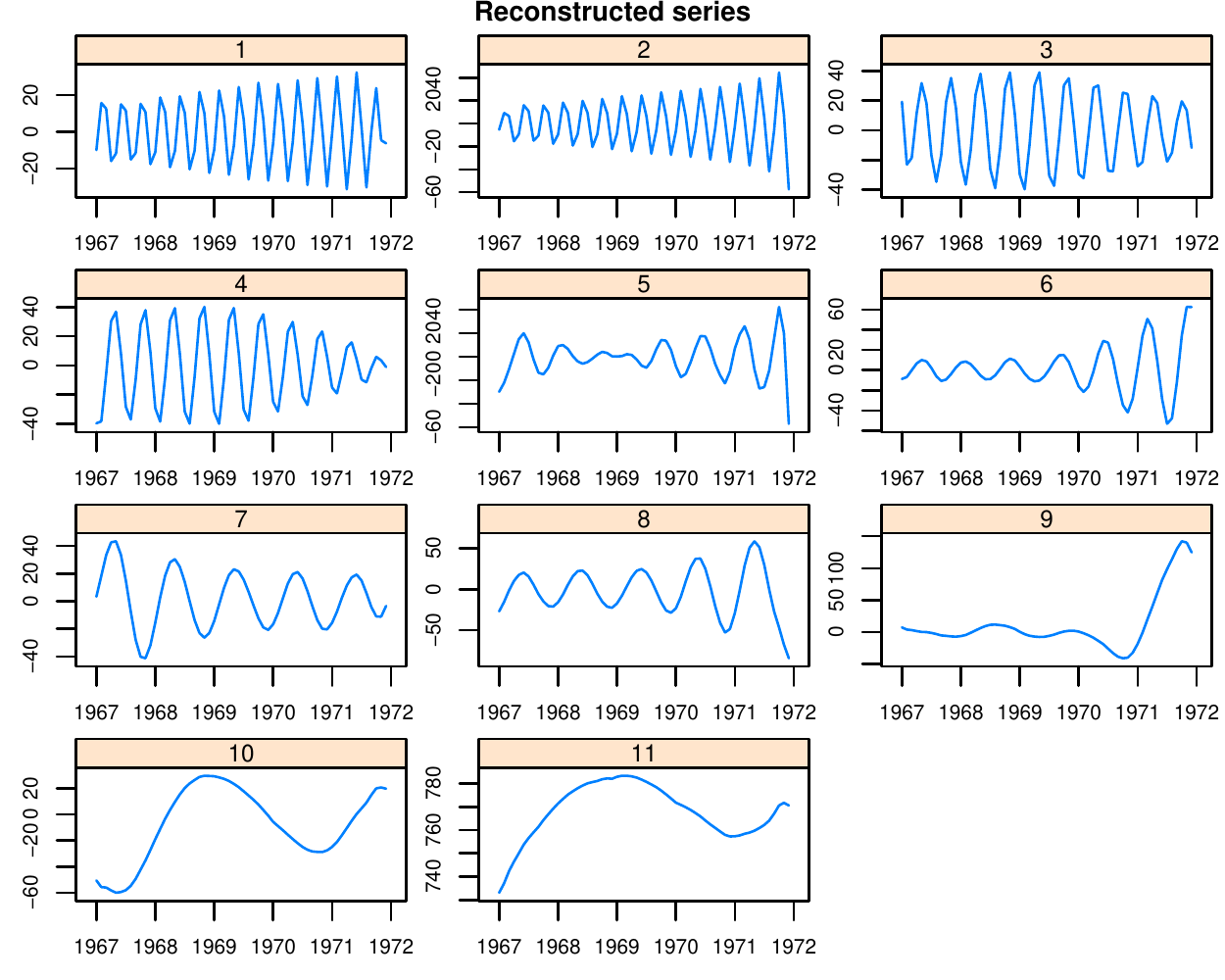}
        \end{center}
        \caption{`MotorVehicle' (5 years), DerivSSA with $L=24$: elementary RCs; good separability.}
        \label{fig:sep_good}
\efg

 \vspace{-0.4cm}
\bfgh
        \begin{center}
        \includegraphics[width=4in]{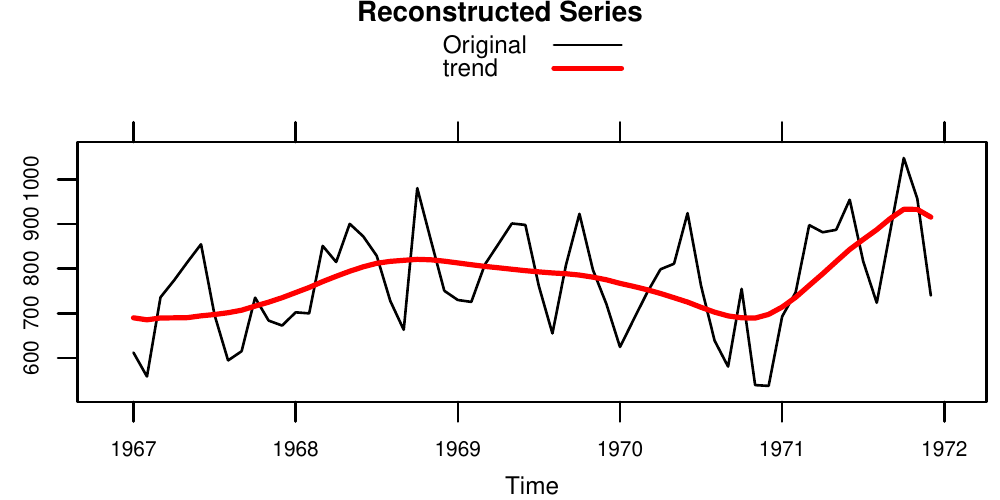}
        \end{center}
        \caption{`MotorVehicle' (5 years), DerivSSA with $L=24$: trend reconstruction, ET9-11; good separability.}
        \label{fig:sep_good_rec}
\efg

\clearpage
\subsection{Filtering}
\label{sec:filtering}

It is known that the time series components reconstructed by SSA
can be considered as a result of linear filters applied to the original time series
\cite{Hansen.Jensen1998,Harris.Yan2010,Bozzo2010} and \cite[Section 3.9]{Golyandina.Zhigljavsky2013}.
Certainly, the coefficients of these filters have nonlinear dependence on the time series.
Therefore, these filters are called adaptive.
This approach is more natural if $L$ is small, since then a larger part of the points of each reconstructed time series is obtained by the same linear filter {(see Section~\ref{sec:SSA_filtering} for details)}.

\subsubsection{Example}
\begin{figure}[!htb]
    \centering
    \includegraphics[width=4in]{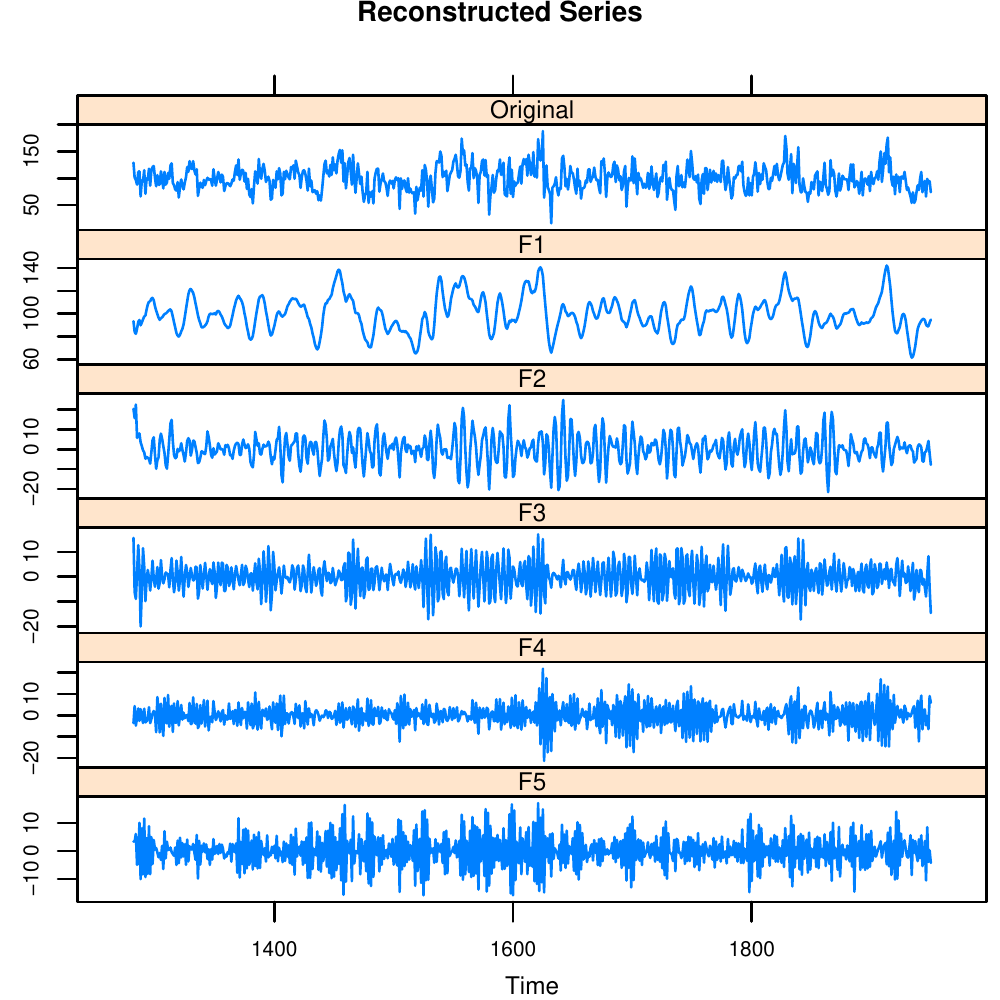}
    \caption{`Tree rings': Frequency decomposition. }
    \label{fig:tree_frequency}
\end{figure}

\begin{figure}[!htb]
    \centering
    \includegraphics[width=4in]{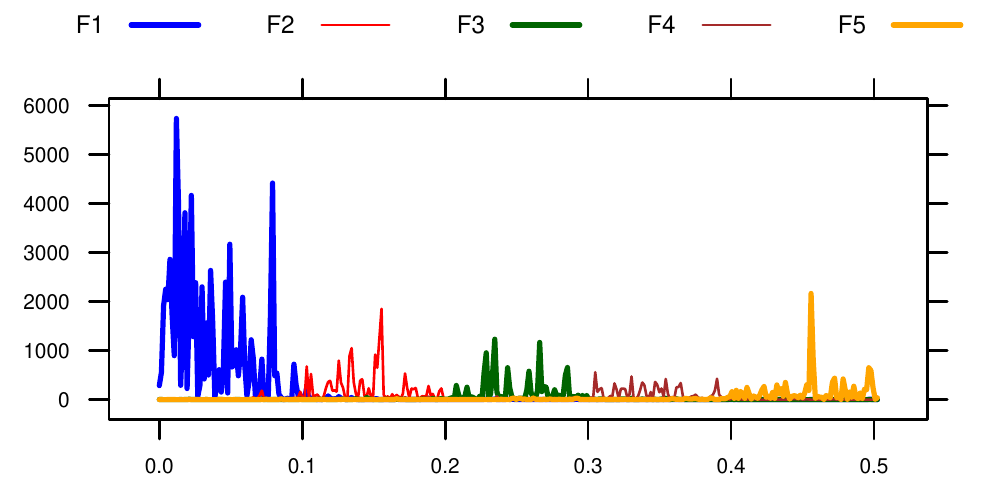}
    \caption{`Tree rings': Periodograms of the reconstructed time series. }
    \label{fig:tree_spectrum}
\end{figure}

The following example shows that SSA can be considered from the viewpoint of
decomposition into components with different frequency ranges, that is, not necessarily
as a decomposition into a trend, periodic components and noise.
Fig.~\ref{fig:tree_frequency} shows the decomposition, which was obtained by
grouping the components according their frequency range.
Fig.~\ref{fig:tree_spectrum} with components' periodograms depicted together confirms this.
This example is performed by the code of Fragment 2.8.1\footnote{\url{https://ssa-with-r-book.github.io/02-chapter2-part2.html\#fragment-281-tree-rings-frequency-decomposition}}
\cite{Golyandina.etal2018}.

\newpage

\clearpage
\subsection{Modelling}
\label{sec:modelling}
We mentioned that SSA combines non-parametric (model-free) and parametric
approaches; certainly, the latter is possible if a parametric model is stated.
Consider the following model of signals that is used in SSA.
Let $\tS = (s_1,\dots,s_{N})$ be a signal (or, more precisely, a time series component
of interest).
Set a window length $L$, $1<L<N$; $K = N - L +1$.
Consider the trajectory matrix:
$$\bfS = \left(\!{\renewcommand{\arraystretch}{1.1}
\begin{array}{@{\ }l@{\;\;}l@{\;\;}l@{\;\;}l@{\;}}
s_1      & s_2 & \dots     & s_{K}     \\
s_2      & \;\adotsss  & \;\adotsss    & s_{K+1}    \\
\; \vdotss   & \;\adotsss   & \ \:\vdotss    & \; \vdotss   \\
s_{L}  & s_{L+1} & \dots & s_{N}
\end{array}
}\right).
$$
Let $r$ denote the rank of $\bfS$.

\medskip
{Different forms of the model} are:
\begin{itemize}
\item
 $\bfS$ is a Hankel low-rank matrix of rank $r<\min(L,K)$; the model can be parameterized by a basis of
 $\colspace(\bfS)$ or of its orthogonal completion. Such time series are called \emph{time series of finite rank}.
 \item
 The time series is governed by a {linear recurrence relation (LRR)}
 \begin{equation}\label{eq:lrf2}
  s_n = \sum_{k=1}^r a_k s_{n-k}, a_r\neq 0, n=r+1,\ldots.
 \end{equation}
 Such time series is called a \emph{time series governed by an LRR}.
 \item
 The time series has an explicit parametric form of a finite sum:
 \begin{equation}\label{eq:model_pol}
 s_n = \sum_j p_j(n)\exp(\alpha_j n) \sin(2\pi \omega_j n + \phi_j),
 \end{equation}
 where $p_j(n)$ is a polynomial in $n$; $\exp(\alpha_j n)=\rho_j^n$ for $\rho_j = e^{\alpha_j}$.
\end{itemize}

The first model is more general; however, under some non-restrictive conditions (for example, if
the time series has infinite length),
these three models are equivalent, see  e.g. \cite[Theorem 3.1.1]{Hall1998} and \cite[Section 5.2]{Golyandina.etal2001}.

Let us describe how the minimal LRR (the irreducible LRR of order that is as low as possible), which governs the time series, determines its explicit parametric form.
A more convenient  form of \eqref{eq:model_pol} is as follows:
\begin{equation}
\label{eq:explicit_complex}
s_n = \suml_{m=1}^p \left(\suml_{l=0}^{k_m-1} c_{ml} n^l\right) \mu_m^n,
\end{equation}
where each $\mu_m$ coincides with $\rho_j e^{\pm\textsf{i}2\pi\omega_j}$ for some $j$.
Thus, the complex parameters $\{\mu_m\}$ in \eqref{eq:explicit_complex} determine frequencies $\{\omega_j\}$ and exponential bases $\{\rho_j\}$ in \eqref{eq:model_pol}.

    The polynomial $P_r(\mu)=\mu^r - \sum_{k=1}^r a_k \mu^{r-k}$ of order $r$ is called
    {\em characteristic polynomial} of the LRR \eqref{eq:lrf2}.
The roots of the characteristic polynomial are called \emph{characteristic roots} of the corresponding LRR.
The roots of the characteristic polynomial of the minimal LRR governing the time series
determine the values of the parameters $\mu_m$ and $k_m$ in \eqref{eq:explicit_complex} as follows.
Let a time series $\tS_{\infty}=(s_1,\ldots,s_n,\ldots)$ satisfy the LRR
\eqref{eq:lrf2}. Consider the characteristic
polynomial of the LRR \eqref{eq:lrf2} and denote its different (complex) roots by
$\mu_1,\ldots,\mu_p$, where $p \leq r$.  All these roots are non-zero as
$a_r\neq 0$ with  $k_m$ being the multiplicity of the root $\mu_m$ ($1\leq
m\leq p$, $k_1+\ldots+k_p=r$). We refer for an extended summary to Section~\ref{sec:parameters} and \cite[Section 2.1.2.2]{Golyandina.etal2018}.

\subsubsection{Subspace-based approach}
\label{sec:subspace_based}
The question is how to find the structure (the basis of $\colspace(\bfS)$, the coefficients of the LRR, the parameters of $s_n$) of a signal $\tS$.
Within SSA, the answer is as follows. Let us apply SSA and obtain the decomposition of the trajectory matrix
$\bfS = \sum_{m=1}^r \sqrt{\lm_m}{U_m}{V_m^\rmT}$.

Then
\begin{itemize}
\item
 The vectors $U_m$, $m=1,\ldots,r$, form a basis of the \emph{signal subspace} $\colspace(\bfS)$.
 \item
 The basis  $U_m$, $m=1,\ldots,r$, provides the coefficients $a_k$ in the LRR $s_n = \sum_{k=1}^r a_k s_{n-k}$, $n=r+1,\ldots$.
 Denote $\pi_m$ the
last coordinate of $U_m$, $\last{U}_m\in \spaceR^{L-1}$ the vector $U_m$ with the  last coordinate removed,
and $\nu^2=\sum_{m=1}^r \pi_m^2$. Then the elements of the vector
\be
\label{eq:minnorm}
R=(a_{L-1},\ldots,a_{1})^\rmT = \frac{1}{1-\nu^2} \sum_{m=1}^r \pi_m \last{U}_m
\ee
are the coefficients of the \emph{min-norm} governing LRR: $s_n=\sum_{k=1}^{L-1} a_k s_{n-k}$
(see discussion of the min-norm LRR in \cite[Section 3.2.3]{Golyandina.Zhigljavsky2013}).
If $L=r+1$, \eqref{eq:minnorm} yields the minimal LRR, which is unique.
{Since LRRs are directly related to forecasting, see also Section~\ref{sec:forecast}.}
 \item
 The basis  $U_m$, $m=1,\ldots,r$, determined the values of $\alpha_j$ and $\omega_j$ in \eqref{eq:model_pol}.
 Consider the signal in the complex-valued form $s_n=\sum_{k=1}^r c_k \mu_k^n$ (we simplify the form excluding polynomials). The relation between parameters is $\alpha = \ln(\Mod(\mu))$ and $\omega = \Arg(\mu)/(2\pi)$. Apply a subspace method, e.g. ESPRIT \cite{Roy.Kailath1989}
 (another name of ESPRIT for time series is HSVD \cite{Barkhuijsen.etal1987}).
 Denote $\bfU_r=[U_{1}:\ldots:U_{r}]$ and let $\last{\bfU_r}$ be the matrix with the
last row removed and $\first{\bfU_r}$ be the matrix with the first row
removed. Then $\mu_k$ can be found as the eigenvalues of the matrix
$\last{\bfU_r^\dag} \first{\bfU_r}$, where $\dag$ denotes pseudo-inversion.
{See Section~\ref{sec:parameters} for more details about the parameter estimation.}
\end{itemize}

This is called \emph{subspace-based approach}. It can be extended to the 2D, 3D, ... cases; certainly, the $n$D theory for $n>1$ is much more complicated than that for the one-dimensional case.

\paragraph{Subspace-based approach in real-world problems}
In real-world problems we observe
$\tX = \tS + \tR$, where $\tS$ is the structured component of interest (e.g. a signal),
$\tR$ is a residual (e.g. noise).\\
Suppose we have an approximate separability of $\tS$ and $\tR$.
Then
\begin{enumerate}
\item
Apply SSA and obtain the set of $U_m$, $m=1\ldots,L$.
\item
Identify the SVD components with numbers $G = \{i_1,\ldots,i_r\}$, which are related to $\tS$; this is possible due to the assumed separability.
\item
Take the set $\{U_i\}_{i\in G}$ as an estimate of the basis of the signal subspace. The same formulas from the subspace-based approach, which are used in the noiseless case, are applied to
$\{U_i\}_{i\in G}$ to get estimates of the LRR coefficients and the time series parameters.
\end{enumerate}

    \subsubsection{Signal extraction via projections}
    \label{sec:signal}
    Consider a particular case of ${\tX}={\tS}+{\tN}$, ${\tX}=(x_1, \ldots, x_{N})$, where $\tS$ is a signal of rank $r$, $\tN$ is noise, and set the parameters: the window length $L$ and
        the signal rank $r$.

    Introduce two projections in Frobenius norm: $\Pi_r: \spaceR^{L\times K} \mapsto \calM_r$, where $\calM_r$
    is the set of matrices of rank not larger than $r$, and $\Pi_\calH: \spaceR^{L\times K} \mapsto \calH$,
    where $\calH$ is the set of Hankel matrices. Let $\calT$ be defined in \eqref{eq:traj}.

     \textbf{Scheme of SSA for signal extraction:}
    \begin{eqnarray*}
        {\tX} \xrightarrow[\fbox{$L$}]{{\cal T}}
        {\bf X} = \left( \begin{smallmatrix}
        x_1 & x_2 & \ldots & x_{K}\\
        x_2 & x_3 & \ldots & x_{K+1}\\
        \vdots & \vdots & \ddots & \vdots\\
        x_{L} & x_{L+1} & \ldots & x_{N}
        \end{smallmatrix} \right)
        \xrightarrow[\fbox{$r$}]{{\rm SVD}: (\sqrt{\lambda_m}, U_m, V_m),\ {\Pi_r}}
    \end{eqnarray*}
    \begin{align*}
        \begin{cases}
        {{\cal L}_r} = \sspan(U_1,\ldots, U_r)\\
        \mbox{is the {signal space}};\\
        \Pi_r\ \mbox{is the projector on}\ {{\cal L}_r};\\
        \widehat{\bf S} = \sum_{m=1}^{r} U_m ({\bf X}^{\rm T} \, U_m)^{\rm T}=\Pi_r {\bf X}.
        \end{cases}
        \xrightarrow{{\Pi_{\cal H}}} \widetilde{\bf S} = \left( \begin{smallmatrix}
        \widetilde{s}_1 & \widetilde{s}_2 & \ldots & \widetilde{s}_{K}\\
        \widetilde{s}_2 & \widetilde{s}_3 & \ldots & \widetilde{s}_{K+1}\\
        \vdots & \vdots & \ddots & \vdots\\
        \widetilde{s}_{L} & \widetilde{s}_{L+1} & \ldots & \widetilde{s}_{N}
        \end{smallmatrix} \right)
        \xrightarrow{{\cal T}^{-1}}
        \widetilde{\tS}.
    \end{align*}
Thus, a concise form of the SSA algorithm for the signal extraction is
$$\widetilde{\tS}=\mathcal{T}^{-1}\Pi_\mathcal{H}\Pi_r\mathcal{T} \tX.$$

\subsubsection{Example of modelling}

Figure~\ref{fig:fort_decomp} depicts the decomposition of the time series `Fortified wines',
which was constructed without the use of a model.
Nevertheless, this decomposition helps to detect the model and estimate its parameters.
The model is a sum of products of polynomials, exponentials and sine waves.
The subspace-based method allows one to construct the parametric model of the signal
of rank $r$.

The subspace-based approach gives us the following form:
\be
\begin{array}{l}
    \nonumber
    \wtilde{s}_n=C_1\, 0.99679^n+C_2\, 0.99409^n\sin(2\pi n/12+\phi_2)+\\  \\ +
    C_3\,1.00036^n \sin(2\pi n/4+\phi_3)+
     C_4\, 1.00435^n\sin(2\pi n/5.97+\phi_4)+\\  \\+
    C_5\, 1.00175^n\sin(2\pi n/2.39+\phi_5)+
    C_6\, 0.98878^n\sin(2\pi n/3.02+\phi_6).
\end{array}
\ee

The coefficients $C_i$ and the phases $\phi_i$ can be estimated by the linear least-squares (LS) method.
{The details are described in Section~\ref{sec:parameters};} the R{}-code for the parameter estimation can be found in Fragments 3.5.9\footnote{\url{https://ssa-with-r-book.github.io/03-chapter3.html\#fragment-359-fort-estimation-of-parameters-by-basic-ssa}} and 3.5.11\footnote{\url{https://ssa-with-r-book.github.io/03-chapter3.html\#fragment-3511-fort-estimation-of-parametric-real-valued-form}} of \cite{Golyandina.etal2018}.

\subsection{Choice of parameters}
\label{sec:choice_param})
The two parameters of SSA are the window length $L$ and the way of grouping.
There are no strict recommendations for their choice. Moreover,
these recommendations differ for different problems and different
assumptions about the time series structure.

For example, if the signal is of finite rank, then $L\approx N/2$ is recommended
(simulations provide the recommendation $L\approx 0.4N$ \cite{Golyandina2010}).

If the signal is not of finite rank or has a complex structure (a large rank and its trajectory
has a large condition number), then a smaller window length is recommended.

If the period of a periodic component of the time series is known, it is recommended
to take the window length divisible by the fundamental period.

For analysis of stationary processes, the recommendation can be special. In particular,
the window length should allow a good estimation of the autocovariance matrix.
This means that $L$ should be small enough.

Formalization of the grouping way is difficult. One approach is based on the separability notion
(e.g. on $w$-correlations as a measure of separability) and on visual inspection of eigentriples
(Section~\ref{sec:ident_rules}).
Another approach can be used in the case of finite-rank signals.
Then the methods of rank detection can be applied (see a brief discussion in paragraph `Signal identification' of Section~\ref{sec:ident}).

\subsection{Theoretical studies}
\label{sec:theory})
Below we touch on several theoretical approaches to SSA.
First, the theory of separability should be mentioned, where separability conditions are
formulated and proved. Recall that the (approximate) separability of time series
components means the ability of the method to construct the decomposition into identifiable
components; in particular, the trend should be separable from the residual to be extracted by SSA
(see Section~\ref{sec:sep_ex}). This theory was developed in \cite{Golyandina.etal2001}
and subsequently different modifications of SSA were proposed to improve separability
(see Section~\ref{sec:nested}).

However, an advanced technique is necessary to obtain theoretical results on the accuracy of reconstruction. A possible approach to this is to use the perturbation technique.
Let us observe a perturbed signal $\tX = (x_1,\ldots,x_N) = \tS + \delta\, \bm\varepsilon$ of length $N$, where the time series $\delta\, \bm\varepsilon$ is considered as a perturbation
of the signal $\tS$. The method results in
$\widetilde{\tS} = \tS + \Delta_N(\tS, \delta, \bm\varepsilon)$; then $\Delta_N(\tS, \delta, \bm\varepsilon)$ is expanded as
$\Delta_N(\tS, \delta, \bm\varepsilon) = \Delta_N^{(1)}(\tS, \bm\varepsilon) \delta + \Delta_N^{(2)}(\tS, \bm\varepsilon) \delta^2+\ldots$.
Most of theoretical results assume
that $\delta$ is small; hence, the first-order error
$\Delta_N^{(1)}(\tS, \bm\varepsilon)$ is studied \cite{Badeau.etal2008, Vlassieva.Golyandina2009, Hassani.etal2011}; this technique is in fact the linearization of the error in the neighbourhood of $\tS$. For simplicity, the behavior of the first order error (as $\delta\rightarrow 0$) is frequently considered as $N\rightarrow \infty$.
This technique allows one to obtain results, which partly help to understand the behaviour of the error.
Nonetheless, this technique is still insufficient, since SSA works for any level of noise
(that is, for any value of $\delta$) as $N\rightarrow \infty$.
In the series of papers \cite{Nekrutkin2010,Ivanova.Nekrutkin2019,Nekrutkin.Vasilinetc2017}, a step ahead in the study of perturbations without the assumption about the smallness of $\delta$ is made; the obtained results are related to the case of non-random perturbation and thereby to the separability of signal components.

It is also worth noting the theoretical approach for the case of a fixed small $L$ and stationary time series,
where $\bfX\bfX^\rmT/K$ tends to the $L\times L$ autocovariance matrix as $N$ (and $K = N-L+1$) tends to infinity (see, e.g., \cite{Huffel1993}). Note that the choice of small $L$ is inappropriate from the viewpoint of separability.

Another strand of theory is devoted to SSA-related methods in terms of the parametric model, low-rank approximations and subspace-based methods (Section~\ref{sec:modelling}).

\subsection{General scheme of SSA decompositions}
\label{sec:general}
We refer to \cite[Section 1.1]{Golyandina.etal2018} for discussion of the general scheme of SSA modifications and variations in detail.
A general form of the SSA-family algorithms for the object decomposition can be presented in the following form.
A wide range of objects, from time series to $n$D shapes, can be considered as the input.

\medskip
\noindent
\textbf{Input}: An object (for example, a time series).
%\medskip
\begin{enumerate}
\item \textbf{Embedding}.
{The input object is transformed into a structured trajectory matrix from a set $\calH$ by an embedding operator $\calT$}
(e.g., $\calH$ is the set of Hankel matrices).
\item \textbf{Decomposition}.
{The trajectory matrix is decomposed into a sum of one-rank elementary matrices}
(e.g., by the SVD).
\item  \textbf{Grouping}.
The elementary matrices are grouped in an appropriate way; the grouped matrices are obtained by summation of
the elementary matrices by the groups.
\item  \textbf{Return to the object decomposition}.
{The grouped matrices are transformed to the form of the input object (e.g., from matrices to time series)
by projecting to $\calH$ (e.g., by hankelization) and performing $\calT^{-1}$.}
\end{enumerate}
%
%\bigskip
\textbf{Output}: Decomposition of the input object into the sum of identifiable objects (e.g., of a trend, oscillations and noise).

\medskip
The SSA algorithm in the general form is easily extended to analyzing objects of different dimensions by the change of the embedding
operator $\calT$. Also, modifications of the decomposition step do not depend on the shape/dimension of the decomposed object,
since they are localized in the decomposition step, whereas the specific of the shape/dimension is localized by means of the embedding operator $\calT$ in the first and last steps.

\subsection{Multivariate/multidimensional extensions}
\label{sec:multi}
Multivariate/multidimensional extensions differ by the embedding step of the SSA scheme;
that is, by definition of the embedding operator $\calT$.
Let us list different versions of the embedding step:
\begin{itemize}
\item
SSA for {time series} (1D-SSA): $\calT(\tX)$ is a Hankel matrix;
\item
MSSA for {systems of time series}: $\calT(\tX)$ is a stacked Hankel matrix;
\item
2D-SSA for {digital images}: $\calT(\tX)$ is a Hankel-block-Hankel matrix;
\item
Shaped SSA for {any shaped objects}: $\calT(\tX)$ is a quasi-Hankel matrix;
\item
M Shaped 3D-SSA for several 3D images of some shapes provides a stacked quasi-Hankel trajectory
matrix;
\end{itemize}
Shaped $n$D-SSA ($n>1$) can be considered as a general $n$-dimensional extension of SSA for shaped objects.

\subsubsection{Multivariate SSA}
\label{sec:MSSA}
Decomposition, forecasting, missing data imputation, and other subspace-based methods are performed in the same manner as for 1D-SSA, see examples from \cite[Chapter 4]{Golyandina.etal2018}\footnote{\url{https://ssa-with-r-book.github.io/04-chapter4.html}}.

Within the framework of SSA, we cannot speak about causality, since the MSSA method is invariant with respect to shifts in time of the time series.
However, we can say about the supportiveness of one series with respect to
another series. The supportiveness of the second time series with respect to the first time series means that the second time series improves the accuracy of signal estimation or forecasting
in comparison with the use of the first series only.
If the time series have a large portion of common structure, their simultaneous processing
is better than the separate processing of each time series. The common structure within the SSA framework means similar signal subspaces.
The accuracy of the signal extraction depends on the signal-noise ratio.
Even if two time series have signals of the same structure, a large noise level in the second time
series can cause this series to be not supportive.

\smallskip
It is important to stress that in MSSA there are a lot of different notations, which are sometimes controversial.
In 1D-SSA, it is not important, what we call left or right singular vectors, eigenvectors or factor vectors/principal components, since the transposed $L$-trajectory matrix
coincides with $K$-trajectory matrix for $K=N-L+1$. Usually, the longer vectors are called factor vectors (principal
components). Another approach is to fix the embedding dimension $L$ (which is equal to the window length $L$) and
to call the right singular vectors factor vectors/principal components. Then increasing the time series length $N$ will increase the length $K$ of factor vectors.
Since usually $L\le K$, both approaches provide the same terminology.

However, for MSSA, this is not the case.
In MSSA, the trajectory matrix is constructed from the stacked trajectory matrices of
time series from the considered collection. The stacking can be either vertical or horizontal.
In the case of horizontal stacking,
$$
\calT_\mathrm{MSSA}(\tX^{(1)},\ldots,\tX^{(s)}) = [\bfX^{(1)}:\ldots:\bfX^{(s)}],
$$
where $\bfX^{(i)}$ is the trajectory (and therefore Hankel) matrix of the $i$th time series.
Therefore, the left singular vectors correspond to time subseries
of length $L$, while the right singular vectors consist of
stacked subseries of different time series from the collection.
Thus, in contrast to 1D-SSA, left and right singular vectors have
different structure.

The vertical stacking is not something different; left and right singular vectors interchange
(after interchanging $L$ and $K=N-L+1$);
therefore, the way of stacking influences no more than terminology, choice of parameters
and possibly computational costs.

{Originally, a small window length $L$ was applied to each of $s$ time series with the vertical stacking of trajectory matrices \cite{Weare.Nasstrom1982}, where $L=3$ (rows of the matrix $\bfX$ were reordered in comparison with the conventional MSSA algorithm with the vertical stacking); therefore, the length $K$ of right singular vectors is large.
The left singular vectors of length $Ls$ were called EEOFs (extended empirical orthogonal functions).
The right singular vectors produce factor vectors and principal components; the latter are equal to the factor vectors multiplied by singular values. Both the factor vectors and principal components have length close to the length of the time series themselves; therefore, they can be used instead of time series in further investigations.

In \cite{Weare.Nasstrom1982}, each time series corresponds to a space point on a 2D surface, that is, $s$ is large.
Each EEOF is divided into three parts corresponding to three time lags (the number of lags is equal to $L=3$) and then each part is depicted as a surface.
In Section~\ref{sec:shaped} we discuss, how to describe the algorithm from \cite{Weare.Nasstrom1982}
as Shaped 3D SSA; it seems this description is more natural.
Note that the algorithm from \cite{Weare.Nasstrom1982} (and all the algorithms which result in EEOFs) corresponds to Decomposition stage and does not contain Reconstruction stage.}

{The vertical stacking was also used in the paper \cite{Broomhead.King1986b}, which is usually considered as
 one of the first papers with the description of MSSA (its Decomposition stage). Although the original trajectory matrices were stacked horizontally in that paper, the authors considered eigenvectors of $\bfX^\rmT\bfX$, not $\bfX\bfX^\rmT$; i.e., the singular value decomposition of the matrix $\bfX^\rmT$ was in fact considered, not of the $\bfX$ itself. To make the difference between the horizontal and vertical variant more understandable, let us note the following: to build the SVD of the trajectory matrix for two time series $\tX^{(1)}$ and $\tX^{(2)}$ with trajectory matrices $\bfX^{(1)}$ and $\bfX^{(2)}$, respectively,
 in the horizontal case, one calculates the eigenvectors of the matrix $\bfX^{(1)}\left(\bfX^{(1)}\right)^\rmT + \bfX^{(2)}\left(\bfX^{(2)}\right)^\rmT$, whereas in the vertical case one finds eigenvectors of the matrix
  $$\left(
\begin{array}{ll}
\bfX^{(1)}\left(\bfX^{(1)}\right)^\rmT &\bfX^{(1)}\left(\bfX^{(2)}\right)^\rmT \\
\bfX^{(2)}\left(\bfX^{(1)}\right)^\rmT &\bfX^{(2)}\left(\bfX^{(2)}\right)^\rmT \\
\end{array}
\right).
$$
}

In \cite{Golyandina.etal2015} and \cite{Golyandina.etal2018}, the horizontal stacking is considered  to fix the number of rows
(that is, to fix the dimension of the column space of the trajectory matrix).
In the case of horizontal stacking, increasing the time series lengths leads to the addition of
columns to the trajectory matrix. Moreover, the lengths of different time series can differ; this does not influence the column dimension.

The above consideration that the horizontal and vertical stackings yield just different forms of the same
method is valid for Basic (M)SSA, where the SVD of the trajectory matrix is considered.
In particular, the form of stacking in MSSA, which was suggested in \cite{Broomhead.King1986b}, does not matter. For Toeplitz (M)SSA, this is not the case. Let $\bfC_i$ be the autocovariance (Toeplitz) matrix
of the $i$th time series, $\bfC_{ij}$ be the cross-covariance matrix
of the $i$th and $j$th time series, $\bfC_{ii} = \bfC_{i}$. Then for the horizontal stacking, the decomposition is constructed
on the basis of the eigenvectors of $\sum_{i=1}^s\bfC_{ii}$.

For the vertical stacking, which is usually considered if MSSA stands for Multichannel SSA,
the decomposition is constructed on the basis of the eigenvectors of
$$\left(
\begin{array}{lll}
\bfC_{11}&\ldots&\bfC_{1s}\\
\ldots&\ldots&\ldots\\
\bfC_{s1}&\ldots&\bfC_{ss}\\
\end{array}
\right).
$$
In climate investigation, the Toeplitz version of MSSA with vertical stacking is conventional \cite{Plaut}.
However, we were unable to find papers where the comparison of the vertical and horizontal versions of Toeplitz MSSA was carried out.

\subsubsection{Two-dimensional SSA}

For digital images given by a 2D array $\tX = \tX_{N_1, N_2}=(x_{ij})_{i,j=1}^{N_1,N_2}$ of size $N_1\times N_2$,
a 2D window size $L_1\times L_2$ should be chosen.
The trajectory matrix consists of vectorized moving 2D windows.
This trajectory matrix can be written in the form of a Hankel-block-Hankel matrix:
\be
\label{eq:block_hankel_matrix}
\bfX = \calT_{\mathrm{2D-SSA}} (\tX) =
\left(\begin{array}{lllll}
\bfH_1        & \bfH_2 \quad      & \bfH_3\quad & \dots  & \bfH_{K_2}   \\
\bfH_2        & \bfH_3            & \bfH_4      & \dots  & \bfH_{K_2+1}     \\
\bfH_3        & \bfH_4            & \!\!\adots& \;\;\adots         & \;\vdots       \\
\;\vdots      & \;\vdots &\!\!\adots &    \;\;\adots      & \;\vdots       \\
\bfH_{L_2}  & \bfH_{L_2+1}\quad   & \dots & \dots         & \bfH_{N_2}
\end{array}
\right),
\ee
where each $\bfH_j$ is the $L_1 \times K_1$ trajectory (Hankel) matrix constructed from
 $\tX_{:,j}$ (the $j$th column of the 2D array $\tX$).

The estimation of frequencies by 2D-ESPRIT is the most frequent use of such trajectory matrix \cite{Sahnoun.etal2017},
see Fragments 5.3.1\footnote{\url{https://ssa-with-r-book.github.io/05-chapter5.html\#fragment-531-mars-parameter-estimation-with-2d-esprit}} and
5.3.2\footnote{\url{https://ssa-with-r-book.github.io/05-chapter5.html\#fragment-532-mars-parameter-estimation-with-shaped-2d-esprit}} from \cite{Golyandina.etal2018}. Also,
2D-SSA is used for the problems of smoothing and noise reduction, see Fragments 5.4.2--5.4.6 \footnote{\url{https://ssa-with-r-book.github.io/05-chapter5.html\#fragment-542-brecon-beacons-decomposition}}.

\subsubsection{Shaped SSA}
\label{sec:shaped}
In \cite{Golyandina.etal2015}, a general approach to singular spectrum analysis
is suggested.
Shaped SSA is the universal version of SSA, which is applicable to arbitrary shapes and dimensions of the objects.
The moving window can also be of arbitrary shape. For example, both the digital image and the moving
window can be circular-shaped; that is, they are not necessary to be rectangular-shaped as in 2D-SSA.
Shaped SSA allows one to consider different versions of the SSA algorithms in a unified manner: SSA for objects with missing data, 1D-SSA, MSSA, 2D-SSA and their circular versions among others. See Fragment 2.6.2\footnote{\url{https://ssa-with-r-book.github.io/01-chapter2-part1.html\#fragment-262-incomplete-decomposition-for-a-series-with-a-gap}} for the 1D case, Fragments 5.2.2--5.2.3\footnote{\url{https://ssa-with-r-book.github.io/05-chapter5.html\#fragments-521-auxiliary-plot-of-2d-image-and-522-mars-mask-specification-and-decomposition}} for the example with a shaped image and a circular-shaped window; and finally Fragment 5.4.6\footnote{\url{https://ssa-with-r-book.github.io/05-chapter5.html\#fragments-546-kruppel-analysis-of-data-given-on-a-cylinder}} for the decomposition of data given on a cylinder \cite{Golyandina.etal2018}.

There is a limitation of this approach, since points within the window and the window locations within the object should be linearly ordered.
Being linearly ordered, the shaped windows are transformed into column vectors and these vectors are stacked to the trajectory matrix according to the ordered locations of the shaped windows.
Many objects can be considered as linearly ordered. The standard technique is to consider the object as a subset of a multidimensional box of the same dimension as the object one, with natural ordering. For example, a piece of the sphere (after its projection to the plane) can be circumscribed by a rectangle; however, there is no continuous planar projection of the whole sphere. Therefore, at the present moment, SSA for data given on the whole sphere is not elaborated.

\subsubsection{Complex SSA}
Complex SSA is Basic SSA applied to complex-valued time series. The only difference in the algorithm is the change of transpose to conjugate transpose.
The algorithm of Complex SSA was explicitly formulated in \cite{Keppenne.Lall1996};
 although SSA was applied to complex time series much earlier, without considering the difference between real-valued and complex-valued data (see, e.g., \cite{Tufts.etal1982}).

Generally speaking, Complex SSA is not a method for analyzing multivariate time series; although, it can be considered
as a special method for the analysis of two time series.

Most applications of Complex SSA are related to the so-called F-xy eigenimage filtering \cite{Trickett2003}. This name is related
to the analysis of digital images in geophysics; first, the discrete Fourier transform (DFT) is applied to each row of the image matrix, then complex-valued series are constructed from the results of the DFTs for each frequency, and finally, the constructed series are analyzed by Complex SSA. Note that the authors of the papers devoted to F-xy eigenimage noise suppression usually omit the word `Complex' in Complex SSA.
The specific of the studied geophysical images is that they are noisy and contain lines (traces); with the help of the DFT, these lines are transformed into a sum of complex exponentials, which have rank 1 in Complex SSA. Therefore, the lines can be separated from noise very well.

\subsection{Modifications of the SVD step}
\label{sec:modif}
Almost independently from the extensions of the embedding step of the SSA algorithm (Section~\ref{sec:multi}), different modifications of the decomposition step
can be considered in a common form $\bfX = \bfX_1 +\ldots + \bfX_r$, where the matrices $\bfX_i$ are some matrices of rank one.

In Basic SSA, the SVD expansion into rank-one matrices is considered.
%It is the best approximating decomposition, which is as well biorthogonal.
If there is no additional information about the time series, the SVD is optimal.
Thus, possible modifications of the decomposition step are related to different assumptions
about the time series.

\subsubsection{Use of a priori information}

The following decompositions have been proposed taking into account the information about the structure of time series:
\begin{itemize}
\item
\textbf{SSA with projection} is used if the model of the signal is partly known; for example, if it is assumed that the signal has a linear trend, see Section~\ref{sec:linear_trend}. We refer to \cite{Golyandina.Shlemov2015} for details, where the general approach with preliminary projections is described. SSA with centering \cite[Section 1.7.1]{Golyandina.etal2001} is a particular case of SSA with projection, since the centering of a vector can be considered as the projection to the subspace spanned by the vector of ones.
 The version of SSA, where rows of the trajectory matrix are centered, came from PCA; however, such centering is not natural for time series, since the rows and columns are subseries of the original time series and therefore have the same structure. Double (both row and column) centering is much more natural for trajectory matrices. It is shown in \cite{Golyandina.Shlemov2015}
that SSA with double centering helps to extract linear trends in the presence of periodic components
(see Fragment 2.8.7\footnote{\url{https://ssa-with-r-book.github.io/02-chapter2-part2.html\#fragment-287-hotel-ssa-with-projection-linear-trend-detection}}).

\item
\textbf{Toeplitz SSA} is used if the time series is stationary (see Section~\ref{sec:stationary}). In the applications related to stationary time series,
Toeplitz SSA is considered as a main version. Details are described in Section~\ref{sec:apriori}.
\end{itemize}

\subsubsection{Refined decompositions of signals}
\label{sec:nested}

Another reason to modify the SVD step is related to improving the separability.
The two main properties of the SVD (the biorthogonality and the optimality) allow Basic SSA to decompose a time series into identifiable components.
The optimality (we mean the approximation properties of the SVD) provides the possibility
to separate signals from noise. The biorthogonality helps to separate time series components if they
are (approximately) orthogonal. It turns out that many time series components such as trends and sine waves
with different frequencies are asymptotically orthogonal (separable) as the time series length tends to infinity; however, we need the separability for time series of finite length. Moreover, the problem of the lack of strong separability (the problem of equal eigenvalues in the SVD) can not be solved even asymptotically.

There are a lot of methods for matrix decompositions, which are not necessarily biorthogonal.
However, most of the methods for a decomposition into rank-one matrices, which are not biorthogonal, simultaneously drop the approximation property.
Therefore, the improvement of separability is performed in two steps. First, the trajectory space
of the signal is estimated by means of grouping the signal elementary matrix components in the SVD step of Basic SSA. Then, the signal grouped matrix is decomposed by a refined expansion (we call such kind of refined expansions nested decompositions).

The examples of SSA modifications with nested decompositions are:
\begin{itemize}
\item
{SSA with derivatives}. This modification can help if the components of the signal were mixed due to equal contributions of
components (that is, in the case, when the weak separability takes place but there is no strong separability) \cite{Golyandina.Shlemov2014}.
The application of SSA with derivatives is demonstrated in
Fragment 2.8.6\footnote{\url{https://ssa-with-r-book.github.io/02-chapter2-part2.html\#fragment-286-us-unemployment-improvement-by-derivssa}}
for the 1D case. Fragment 5.2.5\footnote{\url{https://ssa-with-r-book.github.io/05-chapter5.html\#fragments-524-mars-identification-and-525-mars-improvement-by-derivssa}}
shows that the same approach works in the 2D case.
\item
{Iterated Oblique SSA}. This version helps to solve the problem of
no orthogonality (no weak separability) \cite{Golyandina.Shlemov2014}; see, e.g., Fragment 2.8.2\footnote{\url{https://ssa-with-r-book.github.io/02-chapter2-part2.html\#fragment-282-fort-basic-ssa-and-iterative-o-ssa-trends}}.
\end{itemize}

An approach, which uses additional rotations of mixed components, is also considered
in \cite{Groth.Ghil2011}, where Multichannel SSA with factor rotations is suggested to solve the problem
of the lack of strong separability in MSSA. Recall that this problem is caused by equal eigenvalues in the SVD decomposition
of the trajectory matrix. In the paper \cite{Groth.Ghil2011}, the problem of equal eigenvalues
is called degeneracy of eigenvalues (this is the notion, which is sometimes used in PCA for the explanation of unstable
eigenvectors for close eigenvalues). Formulation of the studied problem in terms of the strong separability may help to gain new insights.

\subsubsection{Tensor SSA}
The general scheme of SSA (Section~\ref{sec:general}) can be further extended by means of considering the embedding operator
$\calT$, which maps the set of objects to the set of tensors of some order instead of the matrix set,
and then using tensor decomposition instead of matrix decomposition in the decomposition step.

The shortcoming is that tensor decompositions are generally not unique and their numerical calculation is time-consuming.
Sometimes, tensor decompositions can be reduced to simpler matrix decompositions.
For example, the SVD of the trajectory matrix in 2D-SSA is in fact the Kronecker-product SVD of
a tensor of fourth order \cite{Golyandina.Usevich2010}.

New algorithms can be obtained by tensor decompositions that are not reduced to matrix ones.
For example, there is a version of SSA where the embedding operator transfers the initial object to a 3D array instead of a matrix (a 2D array); then the tensor decomposition (PARAFAC) is performed,
see, e.g., \cite{Kouchaki.etal2015, Yang.etal2017}.
Note that the idea of the use of tensor decompositions arose much earlier in the subspace-based and
low-rank approximation framework \cite{Papy.etal2005}; see also \cite{Sidiropoulos.etal2017}.

\section{SSA and different problems}
\label{sec:SSAand}
\subsection{SSA and nonlinearity. Is SSA a linear method?}
\label{sec:linear}
In the literature, a criticism of SSA sounds as `SSA is a linear method'.

Let us explain what this means.
First, note that the algorithm is nonlinear. Components, which can be extracted by SSA, are generally nonlinear.
However, the class of time series, which produce rank-deficient trajectory matrices, consists of
time series governed by homogeneous linear recurrence relations (LRRs) in the form $x_n = \sum_{i=1}^r a_i x_{n-i}$.
LRRs are closely related to linear differential equations  (LDEs) $\sum_k b_k s^{(k)}(t) = 0$, where $s^{(k)}(t)$ is the $k$th derivative of $s(t)$, since LRRs
are generated by the finite-difference method applied for solving linear differential equations.
In the theory of dynamical systems, the methods, which are related to linear differential equations, are called linear.
For example, let $s(t) + b s'(t) = 0$, $t\in[0,\infty)$,  be a LDE of order 1; consider $i=1, 2, 3,\ldots$ as the discretization of $t$ and the finite-difference approximation $s'(i)\approx s_{i+1} - s_i$;
then we obtain $s_i + b(s_{i+1}-s_i) = 0$ or, the same, $s_{i+1} = (1-1/b)s_i$, that is, the LRR of order 1.

The answer to the criticism of SSA linearity is as follows.

First, the class of solutions of LDEs is wide enough, since it contains any finite sums
of products of polynomials, exponentials and sine waves.
Then, to deal with signals governed by LRRs (e.g., to extract signals from noisy time series), a large
window length (optimally, in the range from $N/3$ to $N/2$, where $N$ is the time series length) is recommended;
however, smaller windows allow taking into consideration only local finite-rank approximations.
For example, a modulated sinusoid with a slowly-varying amplitude is well approximated by a sinusoid with the same frequency on time intervals of size equal to several periods. Therefore, SSA with a small window length equal to several periods produces reasonable results for this example.
It can be argued that for extraction by SSA, the time series component of interest should be locally governed by the same LRR at each local segment.
Thus, small window lengths allow one to deal with nonlinearity of the model in some common cases, such as modulated harmonics and trends.

A very important feature of SSA is that the method does not use the explicit parametric form \eqref{eq:model_pol} of the time series.
For example, to predict the value of an exponential series $s_n = A e^{\alpha n}$,
one approach is to estimate $A$ and $\alpha$ and then to perform the forecasting by the explicit formula.
However, a more flexible approach is to estimate coefficients of the governing LRR
and to perform forecasting by the estimated LRR. Thus, many SSA-family methods try to avoid the estimation
of parameters in \eqref{eq:model_pol} to make the method more robust with respect to
deviations from the model.

Whereas a slowly changing amplitude of the sinusoidal signal is admissible, a changing frequency is not appropriate for SSA
applied to the whole time series
(except in the case of varying the frequency around one main value; this is called `phase noise').
In the case of changing frequency, methods like EMD + HHT \cite{Huang.Wu2008} can be used for the extraction of oscillations with changing
frequency; however, they only work well if noise is small enough.

\paragraph{Local SSA}
The standard approach to analyzing signals with a changing structure is to consider moving subseries (segments) of the original time series. In the framework of SSA, this procedure is called subspace tracking
(see, e.g., \cite{Badeau.etal2003}). In the papers about subspace tracking, the primary focus
is on the construction of fast algorithms. The other use of subspace tracking is the change-point detection
\cite{Moskvina.Zhigljavsky2003}, \cite[Chapter 3]{Golyandina.etal2001}, see example in Fragment
3.5.12\footnote{\url{https://ssa-with-r-book.github.io/03-chapter3.html\#fragment-3512-sunspots-subspace-tracking}}.

For the estimation of time series components such as trends and signals, SSA applied to moving subseries
  is used (see, e.g., \cite{Leles.etal2018a}). The main problem here is how to combine different local
decompositions into one global decomposition. In \cite{Leles.etal2018a}, central
parts of the local signal reconstructions are used and then stacked. The idea to use only the central part is promising, since
the reconstruction of end points is less accurate. However, the question is why not to use only one central point
(by analogy with the LOESS method \cite{Cleveland1979}); certainly, this can be done if the computational cost is acceptable.

As the result of local SSA, we obtain a signal estimation or just a smoothing.
The problem is how to forecast the extracted signal, since its local estimates
may have different structures on different time intervals. In local versions of SSA,
we do not obtain a common nonlinear model; we have several stacked linear models.
However, many nonparametric local methods including e.g. LOESS
have this drawback.

\paragraph{SSA as linear filter} Another reason to call SSA a linear method is its connection with linear filters
(see Section~\ref{sec:filtering} for references).
However, it should be noted that the coefficients
of the SSA linear filters are produced by the time series in a nonlinear way.

\subsection{SSA and autoregressive processes}
\label{sec:AR}
It is important to remark that the basic model of signals in SSA and
the model of autoregressive processes (AR) are similar only at a superficial glance.
The fact is that the models are totally different.

For SSA, the signal model is $s_n = \sum_{i=1}^r a_i s_{n-i}$ and the observed series is a noisy signal, which has the form $x_n = s_n + \epsilon_n$, where $\epsilon_n$ is typically noise (a non-regular oscillation; for example, a realization of a stationary random process).

For autoregressive processes, we have noise innovations at each step:
$x_n = \sum_{i=1}^r a_i x_{n-i}  + \epsilon_n$.
Under some conditions on the coefficients, such innovations yield that $x_n$, $n=1,2,\ldots$, is a stationary stochastic process, while the series with terms $s_n = \sum_{i=1}^r a_i s_{n-i}$ is a deterministic and not necessarily stationary signal. The coefficients $\{a_i\}_{i=1}^r$ in the SSA model can be  arbitrary.

In both AR and SSA analysis, the characteristic polynomials are constructed on the basis of coefficients $a_i$,
$i=1,\ldots, r$.
In terms of SSA, we are concerned about characteristic roots of the governing LRR, that is, about the roots of the characteristic polynomial
$\mu^r-a_1 \mu^{r-1}-\ldots-a_r$ (see Section~\ref{sec:modelling}).
The roots, which have moduli larger/smaller than 1, correspond to a growing/damped time series components,
whereas roots with unit moduli correspond to a stationary deterministic component like undamped sine waves
(see, e.g., \cite[Example 5.10]{Golyandina.etal2001}).

For the AR model, the AR characteristic polynomial is reciprocal to that in SSA (the coefficients are taken in reverse order); therefore, its roots are inverse to the roots of the characteristic polynomial in SSA with the same coefficients $a_i$, $i=1,\ldots, r$. The stationarity in the AR model corresponds to the case when all the roots
 of the AR characteristic polynomial have moduli larger than 1.
 The case of roots with unit moduli corresponds to non-stationarity
(an example is Brownian motion $x_n = x_{n-1} + \epsilon_n$ and its characteristic polynomial $\mu-1$; the same characteristic polynomial corresponds to a constant signal in SSA).

In the context of SSA used for signal extraction,
AR is mostly considered as a noise model (see, e.g., a discussion of Monte Carlo SSA in Section~\ref{sec:montecarlo}).

There is a common form for forecasts of signals governed by LRRs and for conditional mean forecasts of AR processes.
Both are performed by the LRR with estimated coefficients.
The difference is in the approach for estimating these coefficients.
Note that the forecasting values in the AR model are always converging to zero (or the mean value).

Let us remark that in \cite{Keppenne.Ghil1992} a hybrid of SSA and AR was proposed:
 the AR model was used for forecasting leading elementary reconstructed components obtained by the Toeplitz (VG) version of SSA. In that paper, the proposed forecasting method was applied to climate data, which were modelled as stationary random processes.

\medskip
There is one more connection between SSA and AR.
For autoregressive processes, SSA can be applied to an estimate of the autocovariance function $C(l)$ to find the autoregressive coefficients, since the series $C(l)$ is governed by the LRR with the same coefficients:
$C(l) = \sum_{i=1}^r a_i C(l-i)$  (the Yule-Walker equations).

\subsection{SSA and parameter estimation}
\label{sec:parameters}
In this section, we consider the problem of parameter estimation in the model
described in Section~\ref{sec:modelling}.

Let the signal have the explicit parametric form of a finite sum
\begin{equation}
\label{eq:explicit}
s_n = \sum_j A_j \exp(\alpha_j n) \sin(2\pi \omega_j n + \phi_j)
\end{equation}
and the observed time series be $x_n=s_n+r_n$. Here we simplify the general model \eqref{eq:model_pol}, where
a polynomial multiplier can be presented in each summand.
The complex-valued form of \eqref{eq:explicit} is
\begin{equation}
\label{eq:explicit_complex_nopol}
s_n=\sum_{m=1}^r c_m \mu_m^n,
\end{equation}
$\mu_m=\rho_m e^{\textsf{i}2\pi\omega_m}$.
Here $\omega_m = \Arg(\mu_m)/(2\pi)$, $\rho_m = |\mu_m|$. Recall that  $\mu_m$ are the roots of the characteristic polynomial of
the minimal LRR governing the signal. If $s_n$ are real, then each complex root $\mu_m$
should have its complex conjugate $\mu_k = \mu_m^*$. Therefore, $c_m \mu_m^n + c_k \mu_k^n$ can be made
real by a suitable choice of complex coefficients $c_m$ and $c_k$.

A more general signal model is given by the signal rank: $\rank \bfS = r$, where $r$ is known in advance
or is estimated.
In this model, we consider $s_n = A_1 e^{\alpha_1 n} + A_2 e^{\alpha_2 n}$
and $s_n = A e^{\alpha n} \cos(2\pi\omega n +\phi)$ belonging to the same model for $r=2$, whereas for the explicit representation \eqref{eq:explicit}, these are two different parametric models.

The dependence on the parameters $\alpha_j$ ($\rho_j$) and $\omega_j$ is nonlinear; the dependence
on $A_j$ and $\phi_j$ can be considered as linear, since
$A \cos(2\pi\omega n +\phi) = C_1 \cos(2\pi\omega n) + C_2 \sin(2\pi\omega n)$ for some $C_1$ and $C_2$
depending on $A$ and $\phi$.

The Cram{\'{e}}r-Rao lower bounds (CRLB) for the variance of parameter estimates are known
(see, e.g., \cite{Stoica.Moses2005, Badeau.etal2008a}).
For the case of undamped sinusoids, the CRLB for the estimate of the frequency has order $1/N^3$, while the CRLB for the estimate of the amplitude has order $1/N$.

The common approach for parameter estimation is the nonlinear least-squares method,
which implies the explicit parametric form:
$$
\sum_{n=1}^N (x_n - s_n(\{\alpha_j,\omega_j,A_j,\phi_j\}))^2 \rightarrow \min_{\{\alpha_j,\omega_j,A_j,\phi_j\}}.
$$
It is a very complicated time-consuming optimization problem, since the objective function has many local
minima, the problem is nonlinear and therefore needs iterative methods for solution.

The second approach is called \emph{subspace-based} (Section~\ref{sec:subspace_based}).
Let us consider the complex-valued form \eqref{eq:explicit_complex_nopol}. The nonlinear parameters $\mu_m$
are estimated by one of the subspace-based methods (e.g. ESPRIT), which are based on $U_1,\ldots,U_r$ obtained in Decomposition stage of SSA (Section~\ref{sec:basic_ssa}).
The linear parameters can be found by the conventional linear least-squares method.
Consider the Vandermonde matrix generated by $\mu_m$, $m=1,\ldots,r$:
$$\bfM=\begin{bmatrix}
\mu_1 & \mu_1^2 & \dots & \mu_1^{N}\\
\mu_2 & \mu_2^2 & \dots & \mu_2^{N}\\
\vdots & \vdots & \ddots &\vdots \\
\mu_r & \mu_r^2 & \dots & \mu_r^{N}
\end{bmatrix}.
$$
Then \eqref{eq:explicit_complex_nopol} has the form $\tS = (s_1,\ldots,s_N) = C^\rmT \bfM$, where $C=(c_1,\ldots,c_r)^\rmT$.
If $s_n$ and $\mu_m$ are estimated (the signal $\tS$ is estimated as the reconstructed time series $\widetilde\tS$ obtained by SSA;
$\mu_m$ are estimated as $\widetilde\mu_m$ obtained by ESPRIT), we come to the approximate equality $\widetilde\tS \approx C^\rmT \widetilde\bfM$; then the estimate $\widehat{C}$ can be found by the LS method.

One can see that the described approach is very simple and all we should know about the signal is its rank.
As methods for the frequency estimation, the subspace-based algorithms are called high-resolution, since they provide the frequency estimates with the variance of the same order
$1/N^3$ as the CRLB has; see {\cite{Badeau.etal2008} for the undamped case:
for $s_n=C e^{\textsf{i}2\pi\omega n}$, the ESPRIT estimate of ${\omega}$ has variance $\Var \, \widehat{\omega} \sim 1/N^3$ if the window length $L$ is proportional to the time series length $N$.

\subsection{SSA and structured low-rank approximation (SLRA)}
\label{sec:SLRA}
Let us apply SSA to the problem of extracting a finite-rank signal from a noisy time series.
The algorithm of SSA can be written down in a compact form
by means of projection operators (see Section~\ref{sec:signal}):
$\widetilde{\tS}=\mathcal{T}^{-1}\Pi_\mathcal{H}\Pi_r\mathcal{T} \tX$.

However, in practice, the estimate $\widetilde{\tS}$  is generally not of finite rank.
The problem of finding the estimate $\widetilde{\tS}$ of rank $r$ can be solved
by different methods.
There is a subset of methods, called SLRA, which state the problem of approximation
of the time series trajectory matrix by a low-rank Hankel matrix: $\min_{\bfS \in \calM_r\cap \calH}\|\bfX - \bfS\|_\rmF$.
The most famous method is called Cadzow iterations and was introduced in \cite{Cadzow1988}.
This method consists of alternating projections and can be expressed as
$\tS^{(m)}=\mathcal{T}^{-1}(\Pi_\mathcal{H}\Pi_r)^{m}\mathcal{T} \tX$.
Thus, SSA as a method for signal extraction can be considered as the first iteration of the Cadzow iterations. Note that the Cadzow iterations were suggested in parallel with SSA.

Another approach is to use a particular parameterization of the set $\calD_r$ of time series of rank not larger than $r$.
Then the problem of low-rank approximation
is considered as the least-squares problem
\be
\label{eq:ts_lra}
\min\limits_{\tS\in
\calD_r} \|\tX-\tS\|_\mathrm{w}^2
\ee
 and can be solved by the weighted least-squares (WLS) method.
If the noise is Gaussian, then by the choice of corresponding weights, the weighted LS estimates coincide with maximum likelihood estimates (MLE) and therefore they are asymptotically the best.

The problem \eqref{eq:ts_lra} is still called structured low-rank approximation method
(see \cite{Markovsky2019}), where the Hankel structure of trajectory matrices is considered.
However, the problem \eqref{eq:ts_lra} does not depend on the window length, which determines
the dimensions of the trajectory matrix.
Therefore, in fact, this is not the problem of low-rank matrix approximation \cite{Zvonarev.Golyandina2018}.

Both SSA and Hankel SLRA can be used for signal extraction, forecasting and
frequency estimation, since both methods provide estimates of the signal subspace and
the signal itself.
Comparing the approaches, we can say that
\begin{itemize}
\item {SSA is fast, SLRA is time-consuming;}
\item {SLRA can provide more accurate parameter estimates in comparison with SSA;}
\item {for a signal which only approximately (or locally) satisfies the model, SLRA
does not work; SSA does work;}
\item
{the outcome of SLRA allows simpler
procedures for parameter estimation and forecasting.}
\end{itemize}

Thus, SLRA is applied mostly in signal processing for engineering problems, where the signals
are exactly of finite rank.
 In the general case of real-world time series, the presence of time series components which are exactly of finite rank is a rare case; therefore, SSA is suitable in a much greater extent. For example, slowly-varying trends can be approximated
by time series of finite rank but generally they are not exactly of finite rank.
Also, amplitude modulations of periodic components are not exactly of finite rank for real-world data. However,
this does not create a barrier for SSA to  extract them. Nevertheless, a reasonable example of Cadzow
iterations for extraction of an exponential trend can be found in Fragment
3.5.8\footnote{\url{https://ssa-with-r-book.github.io/03-chapter3.html\#fragment-358-fort-cadzow-iterations}}.

\subsection{SSA and linear regression (LS-estimator of linear trend)}
\label{sec:linear_trend}
As we discussed before, SSA can extract trends, that is, slowly-varying
time series components, since they can be approximated by finite-rank time series.

Ideally (but not necessarily), trends should be series of small rank $r$.
Linear series with terms $s_n = an +b$, $a\neq 0$, $n=1,\ldots,N$, belong to the class of time series, which are governed by LRRs,
and have rank 2. However, linear functions are not natural for SSA
(while exponential series are very natural). The reason is that the characteristic
root for  a linear series is 1 of multiplicity 2 and the minimal linear recurrence formula is
$s_{n+2} = 2s_{n+1} - s_n$ (therefore, the characteristic polynomial is $\mu^2-2\mu+1$). Presence of roots of multiplicity larger than 1 is very unstable.
Any distortion of the coefficients of this LRR transforms the multiple unit root to
two different roots, that is, a linear series to a sum
of two exponentials with small exponential rates or a sinusoid with large period.

There is a modification of SSA, which was called in \cite[Section 1.7.1]{Golyandina.etal2001}
SSA with double centering. In that book, the correspondence between SSA with double
centering and the extraction of linear trends is demonstrated.
As it is shown in \cite{Golyandina.Shlemov2015}, SSA with double centering is a particular
case of SSA with projection, where projections of the rows and columns of the trajectory matrix to given subspaces are produced,
subtracted from the trajectory matrix, and then the SVD expansion of the residual matrix is performed.
SSA with projection is positioned as SSA with the use of some information
given in advance. At the present moment, only the use of SSA with projection for the extraction
of polynomial trends is analyzed.

The most common approach for estimating linear trends is the linear regression method,
where the least-squares solution is used. Let us summarize the results of the comparison
between SSA with double centering and linear regression.
Certainly, if the time series consists of a linear trend and white noise $x_n = an + b +\epsilon_n$,
the least-squares method provides the best estimate of the trend. However, if the residual includes, e.g.,
a periodic component, this is not true.
It is numerically shown in \cite{Golyandina.Shlemov2015} that for the case
$x_n = an + b +\sin(2\pi\omega n + \phi)+\epsilon_n$, where $\epsilon_n$ is white noise,
 the LS method applied to the trend, which was obtained by SSA with double centering, generally overcomes the conventional LS estimate applied to the original time series (i.e., overcomes the ordinary linear regression).

\subsection{SSA and filtering}
\label{sec:SSA_filtering}
Linear finite-impulse response (FIR) filters are defined as $f_n (\tX_\infty) = \sum_{i=-m_1}^{m_2} b_i x_{n-i}$,
where $\tX_\infty = (\ldots, x_{-1}, x_0, x_1, \ldots)$ is the input time series.
The main characteristic of FIR filters is the frequency response $A(\omega)$, which has a simple explanation:
if $x_n=\cos(2\pi\omega n)$, then $f_n(\tX_\infty) = A(\omega) \cos(2\pi\omega n + \phi(\omega))$.

SSA can be considered as a set of adaptive filters (see Section~\ref{sec:filtering} for references).
For the window length $L$, each elementary reconstructed component for the points with numbers from $L$ to $N-L+1$ is obtained
by a linear FIR filter applied to the original time series $\tX=(x_1,\ldots,x_N)$.
If the window length is small, almost all points of the SSA output can be considered as
a result of filtering by an adaptive FIR filter. If $L$ is large, this is not the case.
Actually, each point of the reconstructed series is a linear combination
of values of the original series.
However, for each point with number from $[1,L]$ and $[N-L+1,N]$ we have different linear combinations.

The following result is valid:
let SSA be applied with $L\leq(N+1)/2$ and $U=U_m$ be the $m$th eigenvector in the SVD of the trajectory matrix $\bfX$; $m=1,\ldots,d$.
Then the $m$th elementary reconstructed time series on the interval $[L, N-L+1]$ has the form
\begin{equation}
\label{eq:MPF1}
    \wtilde{x}_n^{(m)}=\sum_{j=-(L-1)}^{L-1}
		\left(\sum_{k=1}^{L-|j|} u_k u_{k+|j|} / L\right) x_{n-j}, \;\; L\le n\le N-L+1.
\end{equation}

This filter is called middle-point filter.
It is shown in \cite[Section 3.9]{Golyandina.Zhigljavsky2013} that
the frequency response $A(\omega)$ of the middle-point filter is determined by the periodogram of $U$.
This explains why we can perform the grouping step based on frequency characteristics of the eigenvectors $U_m$, $m=1,\ldots,d$.

Note that if $U$ consisted of equal numbers, we would obtain the Bartlett (triangle) filter.
The leading eigenvector has vector components close to a constant if the time series is positive and $L$ is small enough (the same sign of the eigenvector components follows from the Perron-Frobenius theorem).

The other helpful property is as follows.
The filter bandwidth tends to be narrower as $L$ (together with $N$) increases.
That is, by increasing the window length $L$ we can obtain a more refined decomposition. Recall that we discuss
the behavior of RCs at $[L, N-L+1]$.

The so-called last-point filter plays an important role in SSA, since it is used for reconstruction of
the last point and therefore related to the prediction of the time series. It is not really a filter, since it is used for reconstructing one point:
$$\wtilde{x}_N^{(m)}=u_L \sum_{i=0}^{L-1} u_{i+1} x_{N-i}.$$
However, it is the only reconstruction filter that is causal (see \cite[Section 3.9.5]{Golyandina.Zhigljavsky2013}
for discussions).

As a rule, causal filters lead to a delay. For example, the causal moving average with the window width $W$
has delay $W/2$. The delay in SSA depends on the separability quality.
If the time series component of interest is exactly separable, then the delay is zero. This is an important advantage of SSA over the moving averaging.
This advantage is a consequence of the adaptive properties of SSA.

\subsection{SSA and ICA}
\label{sec:ICA}

Independent component analysis (ICA) is introduced for random processes
and the word `independent' is related to using
the stochastic independence instead of the uncorrelatedness in PCA that is based on the
orthogonality in the SVD.
In SSA, mostly non-random signals are considered.
Therefore, the direct use of the ICA approach in the SSA algorithm is not appropriate.
In ICA, different measures of stochastic independence are considered.
Once the measure of independence is fixed, the independent components
are found by solving the corresponding maximization problem.
The decomposition step in SSA can be modified in such a manner that the ICA optimization problem is solved instead of the SVD and thereby the
separability of the signal components can be improved. By the reasons explained in Section~\ref{sec:nested},
ICA is used in SSA for a nested decomposition.

 SSA with the SOBI-AMUSE version of ICA
is described in \cite{Golyandina.Lomtev2016}. Note that this version is very similar to the modification of SSA called SSA with derivatives \cite{Golyandina.Shlemov2014},
which was created using a completely different approach.
 The version of SSA with maximization of the entropy is described in \cite[Section 2.5.4]{Golyandina.Zhigljavsky2013}.

Another connection between SSA and ICA
is related to  applications to blind signal separation and is considered in \cite{Pietila2006}.
In that application, Basic SSA is used for pre-processing, i.e. for removal of noise
and for dimension reduction; then ICA is applied for the extraction of independent components
from their mixture in a conventional way. This is similar to the use of PCA before ICA for analysis
of multidimensional data \cite{Kato.etal2006}.

\subsection{SSA and EMD, DFT, DWT}
\label{sec:spectral}
 \paragraph{EMD}
 Empirical mode decomposition (EMD) \cite{Huang.Wu2008} is frequently compared with SSA, since both are model-free techniques.
 It seems that EMD is a method without explicit approximation properties, whereas SSA has both separability and
 approximation properties.
 The first components extracted by EMD is highly oscillating and the last component is referred to a trend; for SSA, it is the opposite, since the signal and the trend typically correspond to the leading components of the decomposition; this is an advantage of SSA as a decomposition method. The advantage of EMD is its ability
 to extract periodic components with complex amplitude and frequency modulations. It is likely that
 the combination of SSA and EMD can extend the range of real-world problems being solved.

 \paragraph{DFT}
 Discrete Fourier transform (DFT) differs from SSA by the use of a fixed basis consisting of sines-cosines with
 frequencies from an equidistant grid against the construction of an adaptive basis in SSA.
  In \cite{Bozzo2010} the relation between SSA and DFT is discussed.
Note that SSA in fact coincides with DFT in the circular version of SSA (see terminology in \cite{Shlemov.Golyandina2014}), where the data are considered given on a circle in the 1D case or on a torus in the 2D case.

 From the viewpoint of frequency estimation (see Section~\ref{sec:parameters}), SSA and the related subspace-based methods  allow one to estimate frequencies with a better resolution than $1/N$, where $N$ is the time series length
 (see, e.g., \cite{SANTAMARIA.etal2000} and \cite{Stoica.Soderstrom1991} for comparisons of DFT, ESPRIT and MUSIC).
 The MUSIC method allows one to construct
 pseudo-spectrums similar to periodograms but with no limitation on the frequencies set.
 Comparing time series models that are suitable for the methods, one can say that a sum of pure sinusoids corresponds to DFT,  while a sum of exponentially modulated sinusoids corresponds to SSA.

 Another application of DFT is the estimation of the spectral density by means of smoothing the periodograms.
 SSA can be used for estimating the spectral density, see, e.g., \cite[Section 6.4.3]{Golyandina.etal2001}, where the results from \cite{GrS58} in terms of SSA are discussed.
 If the spectral density is monotonic, different eigenvectors generally correspond
to different frequencies; otherwise, the eigenvectors are mixed, i.e. they may include
different frequencies with comparable contributions.
 This is an explanation why most of the eigenvectors
produced by white noise (which has a constant spectral density) are irregularly oscillating.
And, vice versa, the eigenvectors generated by red noise (the autoregressive process of order one with a positive coefficient) correspond to distinguishing frequencies.

 In \cite{Yiouetal}, the application of SSA to spectral estimation in climatology
 is discussed.

 \paragraph{DWT} Discrete wavelet transform is the decomposition based on a fixed space-time basis.
 This yields both advantages and disadvantages in comparison with SSA. See discussion in \cite{Yiou.etal2000}.

\subsection{SSA: model-free method and modelling}
\label{sec:SSA_modelling}
In Section~\ref{sec:modelling} we discussed the model of time series
that suits SSA. These are time series of finite rank or, almost the same class, time series governed
by linear recurrence relations. The latter class is slightly narrower; although
it is much more understandable in practice.

SSA is a multi-purpose method, which can be whether model-free or used
for modelling (this is a distinguished feature of SSA). Briefly:
\begin{enumerate}
  \item As an exploratory method, SSA is a model-free technique, which can perform the time series decomposition and the frequency filtering without assumptions given in advance.
  \item If the signal is governed by an LRR, SSA allows one to obtain the explicit form of the signal
  and to estimate the parameters; i.e. SSA is able to perform the parametric modelling.
  \item If the signal satisfies an LRR only approximately (or locally),
  the forecasting and missing data imputation can be performed in the framework of SSA
  without constructing the explicit parametric model; that is, SSA is an adaptable method. This is one of the key advantages of SSA, which
  considerably extends the range of applications.
\end{enumerate}

\subsection{SSA: forecasting and gap filling}
\label{sec:forecast}
In this section, we put the forecasting and gap filling (missing data imputation) together,
since the forecasting can be considered as a particular case of the gap filling
with artificial gaps at the place of the predicted data. On the other hand,
the gap filling can be considered as forecasting internal data.

\paragraph{Parameters finding via cross-validation}
In general, if an algorithm has parameters, their choice can be performed
with the help of the cross-validation procedure, which consists of constructing the prediction on
the training data and then calculation of errors on the test data;
the parameters are chosen to minimize the cross-validation error.

For forecasting, a moving prediction is performed for the cross-validation, where the test
sets step after the training sets.
For gap filling, artificial gaps, which are located at arbitrary
positions, are considered as the test data.
In general, errors of imputations of the test data can be considered for the choice
of forecasting parameters; however, the forecasting accuracy can significantly differ
from the imputation accuracy, since for the imputation we have data from both sides,
while the prediction uses data from one side only and thereby is less stable.
It is an important distinction, since we are looking for a tradeoff
between accuracy and stability, which can be different for prediction and interpolation.

\paragraph{SSA forecasting}
Typically, the model of time series,
which is a sum of a signal of finite rank and noise, is considered for forecasting.
However, the prediction of signal components (e.g. a trend) is possible as well.
Also, the signal or its components should not be exactly of finite rank.
Several subspace-based forecasting methods are suggested in \cite{Golyandina.etal2001, Golyandina.Zhigljavsky2013}:
recurrent, vector and simultaneous ones. The first two versions of the SSA forecasting are
more frequently used. If the signal is exactly of finite rank, all three methods give the same result.

After the decomposition step of the SSA algorithm has been performed, the elementary time series components
can be identified and grouped. Each group consisting of signal components induces a subspace spanned by the eigenvectors from the corresponding group; then the subspace approach ({see Section~\ref{sec:subspace_based}}) is applied to each group separately. The subspace-based methods are fast and work well for forecasting in many cases, when SSA is appropriate. It is convenient that one should not
choose the number of elementary components in advance.

Let us comment the SSA recurrent prediction.
The recurrent forecasting is performed by a special LRR (the min-norm LRR, whose vector of coefficients has a minimal norm), which approximately governs the signal (or the chosen signal component).
This algorithm is known as the linear prediction algorithm \cite{Kumaresan.Tufts1982}, where
the min-norm LRR is used for forecasting.

Let us explain the speciality of the min-norm LRR.
There are a lot of LRRs, which govern the same time series, that is, $s_n = \sum_{i=1}^\tau b_i s_{n-i}$, for various vectors of coefficients $B=(b_1,\ldots,b_\tau)^\rmT$; however, these LRRs have different suppressing properties:
$\Var \hat{s}_n = \Var \sum_{i=1}^\tau b_i (s_{n-i}+\delta_{n-i}) = \|B\|^2 \sigma^2$, where
$\delta_k$, $k=1,\ldots,N$, are white noise,  $\sigma^2 = \Var \delta_k$.
If $\tau=L-1$ is fixed, the SSA forecasting LRR \eqref{eq:minnorm} has the minimum norm of the vector of coefficients and therefore has the best suppressing properties.

Examples of forecasting are shown in Fragment 3.2.1\footnote{\url{https://ssa-with-r-book.github.io/03-chapter3.html\#fragment-321-forecasting-of-co2}}.

\paragraph{Subspace-based missing-data imputation}
The subspace-based approach, which is used for forecasting, can be considered for imputation of missing values as well \cite{Golyandina.Osipov2007}.
Generally, if a prediction algorithm works well for the studied time series, the same algorithm can be
applied to fill a gap by forward prediction from the left,
by backward prediction from the right and then by a combination of the results.

Subspace-based methods work well if the signal subspace can be estimated with sufficient accuracy.
In terms of SSA, this means the necessity of approximate separability of the signal, which we want to analyze, from the residual.

For gap filling, the shaped version of SSA (see Section~\ref{sec:shaped}) is used to obtain the SSA decomposition. Shaped SSA which is applied to time series with gaps limits the choice of the window length, since a sufficient number of windows of the chosen length should be located at the places with no gaps and thereby the separability which needs a large window length can be worsen.
Therefore, the subspace-based methods for gap filling are suitable  only for a small number of gaps; generally, for several compactly located sets of missing values
to allow one to estimate the subspace (Fragment 3.3.1\footnote{\url{https://ssa-with-r-book.github.io/03-chapter3.html\#fragment-331-subspace-based-gap-filling}}).

\paragraph{Iterative gap filling}
In \cite{Kondrashov.Ghil2006}, the approach from \cite{Beckers.Rixen2003} was applied to time series.
The approach suggested in \cite{Beckers.Rixen2003} is very general and can be applied to data in different forms.
The algorithm is iterative and has two parameters (for SSA), the window length $L$ and the rank $r$.
In the first step, the missing data are filled in by some numbers;
e.g., by the average value. Then SSA($L$,$r$) is applied to the obtained time series, which has no missing data, for calculating the reconstructed series.
Next, the values at the positions, which initially were with no missing entries, are changed to the original values; SSA($L$,$r$) is applied to the obtained time series, and so on. Thus, we have repeated iterations. Cross-validation, where artificial gaps serve as the test set, can be used to choose $L$ and $r$.

The iterative method is time consuming; however, this is a universal approach for missing data imputation, since it is applicable
for arbitrary gap locations (see Fragments 3.3.2 and
3.3.3\footnote{\url{https://ssa-with-r-book.github.io/03-chapter3.html\#fragment-332-iterative-gap-filling-one-gap}}).
For forecasting, the iterative subspace-based method is not stable, since it uses the original data, which `hold' the imputation only from the left.

\paragraph{AR and SSA forecasting}
We refer to Section~\ref{sec:AR} for discussion of common and unique features of AR and SSA forecasting.

For the signal model consisting of a trend and seasonality, Seasonal ARIMA can be competitive
with SSA. There are real-world examples where ARIMA provides better accuracy; and vice versa; see, e.g.,
\cite{Hassani.etal2009, DeKlerk2015a}.
One of the advantages of ARIMA is the ability to automatically select the order of the ARIMA model using information criteria
 like
the Akaike information criterion (AIC) or the Bayesian information criterion (BIC).
The number $r$ of signal components can be chosen in SSA also on the basis of the
 AIC/BIC approach; however, as it was discussed in the paragraph devoted to the signal extraction in Section~\ref{sec:ident}, this
 approach has many limitations.
 Note that the Seasonal ARIMA model requires to know the period of the periodic time series component, whereas SSA does not. Moreover,
 observations of a few periods can be sufficient for SSA and definitely is insufficient for
 Seasonal ARIMA.
 Fragment 3.5.18\footnote{\url{https://ssa-with-r-book.github.io/03-chapter3.html\#fragment-3518-sweetwhite--comparison-of-ssa-arima-and-ets}} contains an example of comparison of SSA, Seasonal ARIMA and ETS.

\subsection{SSA and signal detection (Monte Carlo SSA)}
\label{sec:montecarlo}
The problem of signal detection is very important in practice.
If noise is strong, it is easy to find spurious signals, since
noise (being considering as a stationary process) contains all frequencies. The mean contribution of each
frequency is determined by the spectral density of noise. In particular, this implies
that if the spectral density is larger for low frequencies, the probability of
spurious trends or spurious sine waves with low frequencies increases.
This is exactly the case of the so-called red noise (AR(1) with a positive coefficient).

Since we observe one realization of a time series, the contribution
of each frequency from the grid $\{k/N, \ k=0,\ldots,[N/2]\}$ (we consider the periodogram values, which
correspond to these frequencies, as their contributions)
is random with variance, which does not tend to zero as $N$ tends to infinity.
Moreover, it has an exponential distribution, that is, large values are
likely.

The question of existence  of a signal in noise can be reduced to the construction of
a criterion for testing the null hypothesis that the time series is a pure noise;
the criterion should be powerful against the alternative that a signal is present.
There are a lot of such criteria for different models of noise.
Most of them are related to the white noise model.

In the framework of SSA, red noise is the other focus of attention. One of the reasons is
that SSA was primarily popular in climatology, where climatic time series are conventionally
modelled as red noise. In addition, the properties of red noise suit SSA, since
red noise has a monotonic spectral function (see Section~\ref{sec:spectral} for a brief discussion).

The method for detection of a signal in red noise was called Monte Carlo SSA \cite{Allen.Smith1996,Allen.Robertson1996, PALUS.Novotna2004,Jemwa.Aldrich2006,Greco.etal2011,Groth.Ghil2015,Garnot.etal2018}, since
it uses simulations. It seems that the method's name does not reflect its purpose as a method
for hypothesis testing; however, that is the name by which this approach is known.

The approach used in Monte Carlo SSA is straightforward. First, a characteristic of data that reflects the difference between the null and alternative hypotheses should be chosen;
then surrogate data are simulated according to the null-hypothesis to construct the distribution
of the chosen characteristic. Finally, one checks if the chosen characteristic of the real-world data, which can be called test statistic, lies outside $(1-\gamma)/2$-and $(1+\gamma)/2$-quantiles of the constructed distribution. If this is true, the null hypothesis is rejected at the significance level $1-\gamma$. In the case of Monte Carlo SSA, this characteristic is
the squared norm of the projection of the trajectory matrix on a chosen vector which refers to a given frequency. The relation with SSA is in the choice of the projection vector as one of the eigenvectors of the trajectory matrix;
then the test statistic is equal to the corresponding eigenvalue.

Certainly, there are additional problems, which should be solved applying the described scheme.
For example, the parameters of the AR(1) process satisfying the null-hypothesis are unknown and should be estimated.
For signal detection, it is not enough to choose only one characteristic (one projection vector corresponding to one frequency if we speak about Monte Carlo SSA), since a probable signal contain frequencies unknown in advance.  Thus, the problem of multiple testing arises.
We refer to \cite{Golyandina2019} for the description of a more strict statistical approach to constructing the Monte Carlo SSA test, where the problem of multiple testing is solved and an approach for controlling the type I error and estimating the test power is suggested.

\subsection{SSA and outliers}
\label{sec:outliers}
The problem of robustness to outliers is essential for any method. Let us consider
how this problem can be solved in SSA.
Recall that for signal extraction, Basic SSA can be expressed through two projections:
$\widetilde{\tS}=\mathcal{T}^{-1}\Pi_\mathcal{H}\Pi_r\mathcal{T} \tX$ (Section~\ref{sec:signal}).
In Basic SSA, both projections are performed in the Frobenius norm (which can be called the norm in $L_2$).
The squared Frobenius norm is equal to the sum of squared matrix/vector entries.

There are two modifications of SSA, in which the projections are performed in another norm.

\paragraph{Weighted projections} The first approach is to use a weighted norm with different weights of
the time series points, where the points which are suspected to be outliers have smaller weights.
In this approach, the projections are performed in the weighted Frobenius norm.
The weights are chosen by an iterative procedure like that used in the LOWESS nonparametric smoothing
\cite{Cleveland1979} or in the iteratively reweighted least-squares method (IRLS) \cite{Holland.Welsch1977},
where the weights are chosen depending upon the residual values in a specific way.

In this approach, the algorithm of SSA with weights should be implemented. SSA with special weights, where the ordinary SVD
is changed to the oblique SSA, can be implemented with approximately the same computational cost
as Basic SSA \cite{Zvonarev.Golyandina2015}. However, arbitrary weights of different time series points
are not the case. The SVD with arbitrary weights has no closed-form solution and therefore needs an iterative numerical solution.
Thus, the algorithm of SSA with weighted projections, which helps to remove outliers, contains two loops: the inner loop (to calculate the weighted SVD) and the outer loop (to recalculate weights basing on the residuals);
therefore, the algorithm is very time-consuming.
This approach is described in \cite{Trickett.etal2012,Chen.Sacchi2014}, where the authors consider
the weighted projection to the set of low-rank matrices; however, it seems they consider the unweighted projection to the set of Hankel matrices (compare with \cite[Prop.2]{Zvonarev.Golyandina2015}).

\paragraph{$L_1$ projections} The second approach is also frequently used in approximation problems.
To improve the robustness, the projections are constructed in the $L_1$-norm. The idea to use the $L_1$-norm matrix approximation instead of
the $L_2$-norm one (that is, instead of the ordinary SVD if we talk about SSA) is very popular in
data analysis.
Again, $L_1$-SVD has no closed-form solution and therefore time-consuming iterative algorithms should be applied.
There are many papers devoted to $L_1$ low-rank approximations.
 It is shown in \cite{Markopoulos.etal2014} that the optimal solution in the real-valued case has the computational cost of
 order $O(N^{dk-k+1})$, where $d$ is the rank of the data matrix and $k$ is the number of desirable
 components. Therefore, suboptimal solutions are considered to decrease the cost
 (see, e.g., \cite{Kundu.etal2014}).

The $L_1$-projection on the set of Hankel matrices is performed by the change of
diagonal averaging to taking medians instead of averages.
The algorithm with the use of $L_1$-norm is considered in \cite{Kalantari.etal2016}. However,
the problem of its implementation with a reasonable computational cost is still not solved.

\medskip
The previous considerations were related to modifications, which would be robust to outliers.
Another problem is to detect outliers.  A common approach to outlier detection is to
use a change-point detection method; then outliers are removed and the data can be analyzed by a non-modified standard method.
Detection of outliers can be performed by subspace-based methods
with the help of singular spectrum analysis, see \cite[Chapter 3.6.1]{Golyandina.etal2001},
or by standard statistical methods in the SVD step (which is similar to PCA), see, e.g., \cite{DeKlerk2015}.

\subsection{SSA and a priori/a posteriori information}
\label{sec:apriori}
Let us consider what information about the time series can help to modify the SSA algorithm for more accurate estimates
or to analyse the algorithm results.
Note that the general rule is valid: if the used a-priori information is wrong, the modified algorithm can yield
totally wrong results.

The most frequently used a-priori assumption is the stationarity of the time series;
then Toeplitz SSA is used (Section~\ref{sec:stationary}).
In Fragment 2.2.2\footnote{\url{https://ssa-with-r-book.github.io/01-chapter2-part1.html\#fragment-222-simulation-comparison-of-toeplitz-and-basic-ssa}}
 the comparison of
Toeplitz SSA with Basic SSA in dependence on the exponential rate is demonstrated (the rate equal to 0 corresponds to the stationarity).

Another possible a-priori assumption is that the trend is polynomial. Especially for the linear trend,
it is theoretically proved and empirically confirmed that SSA with projection can considerably improve the trend extraction, see Section~\ref{sec:linear_trend} for details.

The second approach can be called posterior \cite{Holmstroem.Launonen2013, Launonen.Holmstroem2017} or bootstrap.
On the basis of the bootstrap approach, bootstrap confidence intervals can be
constructed for any characteristic, which is estimated by SSA; e.g., for the signal itself
or for the signal's parameters. The bootstrap approach includes estimation of the signal and noise
parameters based on the SSA decomposition. Then simulation of a sample consisting of the estimated signal plus
simulated noise allows one to construct confidence and prediction intervals.
Note that the same approach is used in Monte Carlo SSA (which is actually Bootstrap SSA) for testing hypotheses, see Section~\ref{sec:montecarlo}, and in the SSA-forecasting for constructing the bootstrap confidence intervals \cite[Section 3.2.1.5]{Golyandina.etal2018}.

The posterior approach used in \cite{Holmstroem.Launonen2013, Launonen.Holmstroem2017} for the detection of
trend/periodic components tests the stability of the decomposition components to distinguish between
 the signal and noise.

\subsection{SSA: automatic identification and batch processing}
\label{sec:ident}
Let us describe approaches to the automatic identification of eigentriples in SSA for their grouping and then extracting the trend and periodic time series components.

\paragraph{Trend identification} The methods proposed for the automatic identification of the trend components are quite natural, since the trend can be described in a nonparametric way as a low-frequency component of the time series.

In \cite{Vautard.etal1992}, different methods for the detection of trend components in the SVD of the trajectory
matrix were suggested; in particular,
the number of zeros or the Kendall's tau correlation coefficient were considered for the detection of trend
eigenvectors in the grouping step of the SSA algorithm. The number of zeros shows (in an indirect way) if a component is low-frequency. The Kendall's tau correlation reflects if a component is increasing or decreasing.

In \cite{Alexandrov2009} and \cite[Section 2.4.5.2]{Golyandina.Zhigljavsky2013}, low-frequency components
are extracted in a direct way by analysis of component's periodograms. More precisely, eigenvectors
or factor vectors (or elementary reconstructed time series) taken from the SSA decomposition are considered. Then a frequency range $[0,\omega_0]$  and a threshold are chosen. If the contribution of frequencies from the
given frequency range is larger than the threshold, the component is referred to as the trend one.
This simple algorithm works very well if the trend components are separated from the residual.
If the trend does not have a complex form, the trend is usually well separated, see Fragment
2.8.9\footnote{\url{https://ssa-with-r-book.github.io/02-chapter2-part2.html\#fragment-289-paynsa-automatically-identified-trend}}.
A slightly different approach is described in  \cite{Watson2016}.

\paragraph{Periodicity identification} The approach based on component's periodograms can be extended to detecting harmonics (sine waves).
A specific feature of the harmonics extraction is that a sine wave produces two components in the SSA decomposition
for any frequency in $(0,0.5)$ and only one component for the frequency 0.5.
The algorithm of the recognition of paired sine-wave components based on the component's periodograms was suggested in \cite{Vautard.etal1992}
and studied for the application to exponentially-modulated harmonics in \cite{Alexandrov.Golyandina2005}.

Whereas the trend is  as a rule well separated from the residual, pairs of components produced by different harmonics
can mix if the harmonic's amplitudes are close. Two SSA modifications, Iterative Oblique SSA and SSA with
 derivatives (see Section \ref{sec:nested}) can be applied for improving the separability
 before the use of identification algorithms.

\paragraph{Grouping} Above, we described different approaches to the trend identification and the identification of periodic components.
A common approach to the automatic grouping is to apply a clustering algorithm to the matrix of weighted correlations between the elementary reconstructed time series.
If the time series components are well separated, this approach works well,
see Fragment 2.7.1\footnote{\url{https://ssa-with-r-book.github.io/01-chapter2-part1.html\#fragment-271-white-dwarf-auto-grouping-by-clustering}}.
However, this way of grouping fails if the groups are poorly separated.

\paragraph{Use of automatic identification} Automatic identification and batch processing have their own parameters, which can be chosen
according to the assumed structure of the time series components of interest. Therefore, these techniques
work in the case of analysis of a collection of similar time series. Generally, to choose parameters of the identification procedure, a preliminary analysis of several time series should be performed in an interactive manner.

Note that a method of automatic identification, which calculates some characteristics of the decomposition components and then compares
them with a threshold, can provide a helpful guess for the interactive grouping based on the values
of the considered characteristics.

\paragraph{Signal identification} A completely different way of identification of the signal components is based on the approach that uses a parametric model.
If the signal is assumed to be of finite rank $r$ and is dominated, that is, the $r$ leading
components correspond to the signal in the SSA decomposition, then the model (which is determined by the signal rank)
can be chosen by information criteria like AIC or BIC (BIC is recommended).
The information criteria need finding the MLE of the model parameters, whereas, as a rule, an LS estimate of the signal is constructed
within low-rank approximation methods (Section \ref{sec:SLRA}). The WLS estimate with appropriate weights, which coincides with the MLE if the residuals are Gaussian, is considered in \cite{Zvonarev.Golyandina2018}, where
  a fast algorithm for the WLS estimation is proposed;
in particular, the difference with the approach from \cite{Usevich.Markovsky2014} is discussed.

Consider the simple case of white Gaussian noise, where the ordinary LS estimate and the MLE are the same.
Denote $\wtilde\tS(d)$ the LS estimate of the signal of length $N$ assuming the parametric model of time series
of rank $d$.
The number of parameters is $2d$.
Define $\mathrm{RSS}(d)=\|\wtilde\tS(d)-\tS\|^2$.
Then, by definition, $$\mathrm{AIC}(d) = N\ln(\mathrm{RSS}(d)/N) + 4d, \qquad
\mathrm{BIC}(d) = N\ln(\mathrm{RSS}(d)/N) + 2d\ln N.$$
The values of AIC/BIC can be used for the choice of rank $r$ in a conventional way:
the estimate of $r$ is the minimum point of the considered information criterion.

Information criteria `as is' can be used in the SLRA statement of the problem, see
Section~\ref{sec:SLRA}, if the signal is of finite rank and we are able to
construct its MLE.
In practice, signals are only approximated by time series of finite rank.
Even if the signal is of finite rank, the signal estimate given by SSA is generally not of finite rank
and is not the LS estimate.
Thus, the use of information criteria for the rank estimation is questionable in their application to
real-world problems.

The other general approach applied to choosing the rank $r$ automatically is the cross-validation, which is briefly discussed in Section~\ref{sec:forecast}. This approach is time-consuming and can be applied
to only long time series; moreover, the aim of this approach is to find $r$ for better
prediction/gap filling, not for signal rank estimation. However, this technique is appropriate under much weaker
assumptions about the signal and noise; therefore, it is applicable in practice.
The \textsf{R} code for the choice of the signal rank $r$ by cross-validation can be found
in Fragments
3.5.13--3.5.15\footnote{\url{https://ssa-with-r-book.github.io/03-chapter3.html\#fragment-3513-functions-for-the-search-of-optimal-parameters}}.

\subsection{SSA and machine learning}
\label{sec:ML}
As we have mentioned, SSA can be called principal component analysis (PCA) for time series. Therefore, the use of SSA in many cases
is similar to the use of PCA (SVD) for multivariate data. In \cite[Section 1.7.3]{Golyandina.etal2018}
one can find a brief review of papers, where SSA together with some other methods (SVM, SVR and NN among others) are used in machine learning.

\section{Implementation of SSA}
\label{Rssa}
\subsection{Software and fast implementation}
\label{sec:software}
At the moment, there are a lot of different implementations of SSA. Let us enumerate
several of them:
\begin{enumerate}
    \item
    the general-purpose interactive {`Caterpillar'-SSA software} (\url{http://gistatgroup.com}, Windows);
    \item
    the interactive software oriented mainly on climatic applications, SSA-MTM Toolkit for spectral
    analysis and its commercial extension kSpectra
    Toolkit (\url{http://www.atmos.ucla.edu/tcd/ssa}, Unix, Mac);
    \item the commercial statistical software SAS, which includes SSA to its
    econometric extension SAS/ETS\circledR;
    \item
     the \textsf{R} package \textsf{Rssa}, a cross-platform implementation of
    a lot of SSA-related methods (\url{http://cran.r-project.org/web/packages/Rssa}).
\end{enumerate}

Fast effective algorithms are implemented in the \textsf{Rssa} package,
where the computational cost (in flops) is dropped from  {$O(N^{3})$} down to {$O(k N \log{N} + k^{2}N)$}}
and the memory consumption is reduced from {$O(N^{2})$} to {$O(N)$};
here $N$ is the time series length, $k$ is the number of calculated eigentriples, the window length $L$
is considered proportional to $N$.
Briefly, the approach is based on the Lanczos algorithm and on  computing the vector
multiplication through Fast Fourier Transform applied for the calculation of convolutions
\cite{Korobeynikov2010,Golyandina.etal2015}.

Note that the \textsf{Rssa}-implementation of the SSA decomposition does not take into consideration if the data were updated (if new data were appended to the time series).
Therefore, the application of SSA to the updated data doubles the computational cost.
There are different approaches to updating the SVD.
However, it is still an unanswered question whether an algorithm for updating SSA can be faster than
the current implementation of SSA.

Let us finally remark that reasoning about the computational cost of SSA-related methods, which are reported in the literature, can be irrelevant; e.g., the sources can use only the information known on the publication date.
Let us give some  examples.
In many papers, SSA is considered as a very time-consuming method because of the used SVD expansion; in particular,
in some past papers, the method ESPRIT is called time-consuming (with computational cost {$O(N^{3})$} as the time series length tends to infinity). However, the method
implementation in \textsf{Rssa} is much faster.

Another example is the SSA vector forecasting. In
\cite{Golyandina.etal2001}, this method is called very time-consuming in comparison with the SSA recurrent forecasting.
In \textsf{Rssa}, the implementation of the SSA vector forecasting \cite{Golyandina.etal2015} is even faster than that
of the recurrent one. On the other hand, this does not mean that the recurrent forecasting cannot be done faster
in next implementations.

One more example is related to mathematical issues. In SSA and especially in MSSA, the understanding that for the original trajectory matrix and for the transposed one the SVD decompositions are in fact the same can help to considerably decrease the computational cost by choosing the case, which is less time-consuming for the used numerical algorithm.

\subsection{Example of calculations in \textsf{Rssa}}
\label{sec:ex_rssa}
Let us demonstrate how fast are the computations in \textsf{Rssa}.
For the time series length $N=1000000$ and the window length $L=500000$,
the reconstruction of a sine wave signal based on two leading components is executed in a few seconds:
\small
\begin{verbatim}
> library("Rssa")
> N <- 1000000
> signal <- sin((1:N)*2*pi/10)
> ts <- signal + 10*rnorm(1:N)
> system.time(s <- ssa(ts, L = N/2, svd.method = "auto", neig = 2))
        user       system      elapsed
        1.19         0.16         1.34
> system.time(rec <- reconstruct(s, groups = list(sig = 1:2)))
        user       system      elapsed
        0.55         0.13         0.67
> max(abs(signal - rec$sig))
[1] 0.0515102
\end{verbatim}

\section{Conclusion}
\label{sec:concl}
As the readers can see, even a brief description of SSA-related themes composes a very large paper.
It is therefore difficult to complete this review paper with a concise conclusion.
Summing up, we want to express the hope that the paper can help researchers from various scientific fields to gain new insights and successfully apply SSA to their studies together with other standard methods.

\section{Acknowledgement}
I am grateful to my co-authors of the papers and monographs devoted to SSA for the joint work and fruitful discussions, which gave me the possibility to look at SSA from different points of view.
My special thanks to Vladimir Nekrutkin for the provided review of the SSA literature of the past years.

\end{document}